\documentclass[preprint2]{aastex}

\slugcomment{Not to appear in Nonlearned J., 45.}

\shorttitle{Flare events on 20 November 2003}
\shortauthors{Kumar et al.}

\begin{document}


\title{Evolution of Solar Magnetic Field and Associated Multi-wavelength
Phenomena: Flare events on 20 November 2003}


\author{Pankaj Kumar}
\affil{ Aryabhatta Research Institute of Observational Sciences (ARIES), Manora Peak, Nainital-263 129, India}
\email{pkumar@aries.res.in}

\author{P. K. Manoharan}
\affil{Radio Astronomy Centre, Tata Institute of Fundamental Research, Udhagamandalam (Ooty)-643 001, India}
\email{mano@rac.tifr.res.in}

\and

\author{Wahab Uddin}
\affil{Aryabhatta Research Institute of Observational Sciences (ARIES), Manora Peak, Nainital-263 129, India}
\email{wahab@aries.res.in}



\begin{abstract}
We analyse H$\alpha$ images, soft X-ray profiles, magnetograms, extreme ultra-violet images and, radio observations of two homologous flare events (M1.4/1N and M9.6/2B) on 20 November 2003 in the active region NOAA 10501 and study properties of reconnection between twisted filament systems, energy release and associated launch of coronal mass ejections (CMEs). During both events twisted filaments observed in H$\alpha$ approached each other and initiated the flare processes. However, the second event showed the formation of cusp as the filaments interacted. The rotation of sunspots of opposite polarities, inferred from the magnetograms likely powered the twisted filaments and injection of helicity. Along the current sheet between these two opposite polarity sunspots, the shear was maximum, which could have caused the twist in the filament. At the time of interaction between filaments, the reconnection took place and flare emission in thermal and non-thermal energy ranges attained the maximum. The radio signatures revealed the opening of field lines resulting from the reconnection. The H$\alpha$ images and radio data provide the inflow speed leading to reconnection and the scale size of particle acceleration region. The first event produced a narrow and slow CME, whereas the later one was associated with a fast full halo CME. The halo CME signatures observed between Sun and Earth using white-light and scintillation images and in-situ measurements indicated the magnetic energy utilized in the expansion and propagation. The magnetic cloud signature at the Earth confirmed the flux rope ejected at the time of filament interaction and reconnection.

\end{abstract}


\keywords{Magnetic reconnection, Solar Flare, Coronal Mass Ejections, Filaments}




\section{Introduction}
It is now generally accepted that the stressed magnetic energy stored in the magnetically complex active region is the source of flare energy, which is seen in a wide spectrum of energy band. The transfer of magnetic energy takes place at the reconnection site, where the localised dissipation region is formed at the current sheet and the magnetic energy is converted into thermal energy and utilized for bulk acceleration of plasma and also for changing the topology of the field lines. The total magnetic energy dissipated could roughly divide in equal parts of thermal energy and kinetic energy (e.g., Priest and Forbes 2000).

In the case of an eruptive flare, which is known to be associated with the launch of coronal mass ejection (CME), the mechanical energy and the dissipative reconnection of the unstable MHD system can give rise to the ejection of flux rope. Observations of CMEs in the interplanetary medium confirm the ejection of `fluxrope' in the form of `magnetic cloud' (e.g., Marubashi 1986, Burlaga et al. 1998). A huge quantity of coronagraph data has established that a fairly large CME can destabilize the large part of the corona and its extension is of the order of one solar radius in the near-Sun region. Thus a CME goes through considerable expansion and as it propagates to farther heights, its size evolution and the propagation speed in the inner heliosphere shows the transfer of magnetic energy to the background solar wind (e.g., Manoharan 2006) in the form of aerodynamical drag force.

The physical conditions to destabilize the coronal structure and to initiate a CME have been explained by several authors (e.g., D{\'e}moulin and Berger 2003, Low 2001). Moreover, the filament interaction followed by the reconnection and resulting eruption has also been studied (e.g., Su et al. 2007). For example, the soft X-ray images from Yohkoh satellite have provided several examples of eruption of twisted coronal loops (S or inverse-S shaped) (i.e., Manoharan et al. 1996, Pevtsov et al. 1996). Such eruptions can be triggered by the loss of equilibrium or rapid emergence of new flux in and around the active region (e.g., Kurokawa 1987a, Kurokawa et al. 1987b). A cusp formation, i.e. twisted magnetic field lines rise to a considerable height, and its eruption may lead to the restructuring of the magnetic configuration. In the highly conductive corona, the dissipation of magnetic energy associated with the reconnection can be rather fast. However, the formation of a cusp and its eruption is important because it involves fast rate of reconnection as well as ejection of helicity (i.e., twisted field) into the interplanetary space. The next natural question is that how the helicity is accumulated, which leads to the formation of twisted-cusp shape. Recently several authors have reported that the sunspot rotation can lead to the building of magnetic energy, i.e., addition of helicity by emerging flux (i.e., Low 1996, Zhang and Low 2005, Zhang et al. 2008, Yan et al. 2008, Chandra et al. 2009)

The period of October-November 2003, `Halloween Days', is well known for extreme level of
solar activities, corresponding to active regions NOAA 10484, 10486 \& 10488. 
The active region NOAA 10501 (return of NOAA 10484, from the previous rotation) showed 
continuous emergence of magnetic flux in and around the activity site and 
produced 17 M-class flares (e.g., Gopalswamy et al. 2005, Chandra et al. 2009). Some of these Halloween-day events provided opportunities to understand the destabilization of magnetic field, formation of cusp, and associated flares and CMEs.
In this paper, we report the rotation of sunspots, leading to the formation of cusp followed by reconnection, initiation of eruptive flares and associated CMEs in the interplanetary space. We study two homologous flare events (M1.4/1N, M9.6/2B) on 20 November in the active region 10501. During these events, the magnetic configuration at the activity site (located $\sim$N00 W05) was rather complex. The combined observations from ground-based H$\alpha$ and radio measurements and space-borne white-light, EUV images, and magnetograms provided opportunities to study these flares and CME events and their coronal structures. The interplanetary consequences of these events are studied based on white light, scintillation images and in-situ solar wind data at the near-Earth environment. The Sun-Earth connections obtained from this study show that studies of such events are essential to outline the space-weather effects of Earth-directed solar events.


\section{Observations}

Figure 1 displays the Michelson Doppler Imager (MDI) magnetogram and white-light image of AR 10501 observed on 20 November 2003. As shown by the magnetogram, the magnetic configuration of the active region was rather complex. On the above date, the active region was located close to the centre of disk, $\sim$N00 W05. More discussions on MDI images are given in Section 2.2. Figure 2 (top panel) shows the soft X-ray flux recorded by GOES satellite in the 0.5-4 and 1-8 \AA \ wavelength bands during 01--09 UT.  The first flare event (M1.4) started $\sim$ 01:45 UT, attained a maximum $\sim$02:05 UT and then gradually decreased. As seen in the plot, the rise as well as the fall in the X-ray flux is rather gradual and the profile is broad with an effective width of $\sim$30 minute or base width of $\sim$1 hour. This suggests a gradually evolving magnetic field and associated energy release. The second event starts with an impulsive weak flare (C3.8) at 07:30 UT. The X-ray profile, unlike the previous one rises quickly to the maximum at $\sim$07:45 UT. It is however evident that the area under the curve of the first flare is equivalent to that of the second event. In this figure, the X-ray peak seen in between the above events corresponds to an activity at a different active region.  


\subsection{H$\alpha$ Observations}

The H$\alpha$ observations of these flares have been carried out at ARIES, Nainital, India, by using 15 cm f/15 coud\` e solar tower telescope. At ARIES, the images are enlarged by a factor of two using the Barlow lens and each image is recorded by a 16-bit 576$\times$384 pixels CCD camera system having pixel size 22 ${\mu}$${m^2}$. The resolution on the image is $\sim$1\arcsec per pixel and the cadence is $\sim$15-20 sec per image. On 20 November, the H$\alpha$ data covered the time interval 01:25 to 08:25 UT. The relative H$\alpha$ intensity profiles of these two flare events are shown in Figure 2 and they look nearly similar to that of soft X-ray flux profiles.

The H$\alpha$ profile corresponding to the first event however shows a flash phase at the start of the flare, whereas the later event shows multiple peaks over the flaring period. It is evident that as seen in X-ray data, the equivalent width of the H$\alpha$ emission profile is nearly twice for the first event, compared to that of the second one. Figure 3 shows the evolution of the first flare (1N/M1.4) in H$\alpha$, during 01:27--02:17 UT. The active region showed several curved and twisted filaments. But the magnetic field structures evolved gradually over the period of the flare. For example, two twisted filaments (indicated by arrows) showed structural and positional changes as the flare was developing over the period of $\sim$ 25 minutes. The filament in the east approaches towards the filament at the west, close to the centre of the active region. But as it moves, the eastern one curves at the middle portion, where the distance between the filaments decreases. When the interaction is effective at about 01:55 UT, the reconnection takes place at the X point, we notice filaments moving to north and south directions. Figure 4 shows the typical cartoon for the above scenario: (a) and (b)show approaching filaments, (c) displays the interaction and resulting reconnection and (d) shows the resulting north and south moving filaments. In Figure 5, the observed distance between the two filament systems has been plotted as a function of time. As the distance decreases, the flare maximizes and the `distance-time' profile provides a relative sky-plane speed of $\sim$10 km s$^{-1}$. Thus, the interaction of the filament systems has likely triggered the flare and it is consistent with the flare start at $\sim$01:45 UT and maximum at $\sim$01:52 UT. The interaction, merging and reconnection are shown by images taken between 01:27--01:51 UT. At the time of merging of filaments, the rearrangement of field lines has also been observed, leading to the reorientation of the filament system. It shows that the occurrence of the flare is due to the motion and the destabilization in the filament system. After the flare, the filament systems get restructured and stabilized. The following sequences:  (1) approaching filaments (2) interaction (3) flare onset (4) separation of filaments in different direction (above and below the X-point) after reconnection suggest interlinked sequence of events in generating the CME. The approach speed of the filaments is in agreement with the earlier inflow speed by X-ray measurements (e.g., Yokoyama et al. 2001).

During 02:40 -- 07:30 UT, there was no significant activity at the flare
site (refer to Figure 1). It is likely that the energy building has been going
on in this time interval. As in the above flare, during the onset of 
the second event, $\sim$07:30 UT, the twisted filament systems, one foot 
point attached to the sunspot group, goes through heavy destabilization
and filaments approach each other. But, the speed of approach has been considerably faster than the previous case. In Figure 6, sequence of selected H$\alpha$ images are displayed for the second flare interval, for which the intensity is significantly high during the flare maximum (also refer to Figure 2). At the time of flare increase, we see an arch like brightening which expands towards north-east of the flare centre. It suggests that the reconnection process between the 
twisted field lines and nearby small-scale field lines. Further,
as the merging of the filaments has been progressing (refer to Figure 6), the
flare raises to the maximum of intensity (refer to Figure 1 and 2). The restructuring  
of filaments after the maximum has been clearly seen in H$\alpha$. The magnetic structure at 07:40 UT is associated with the pumping of the stored magnetic energy which is released during the main phase of the flare and it is in good agreement with the sudden jump in soft X-ray flux. During the flare maximum, 07:45--07:50 UT, at the top of the dark cusp shows the opening of the field lines (refer to section 2.4). 

The dark cusp-shaped structure formed during the flare (refer to Figure 6) starts to move upward. Its speed in the plane of disk is about $\sim$75--100 km s$^{-1}$, during 07:45 - 07:50 UT. In the sequence of images, at 07:51 UT the cusp shape erupts and some of its material also falls back. The cusp speed suggests the possible initial speed of the CME, $\sim$100 km s$^{-1}$, at the low coronal heights. More interesting point is that the cusp motion correlates with the gradual increase in the width of H$\alpha$ flaring region, at the centre of the active region. It is likely to be associated with evolution of the flare ribbons. The rate of widening of flare ribbons excellently correlates with the rising rate of the erupting height of cusp. Figure 7 shows the plot  of ribbons separation and cusp expansion plotted against time. The high degree of correlation between their rates ($\sim$93\%) suggests that the magnetic reconnection and the rising of filaments system have played a prime role in the formation of the CME. The interesting point is that the projected speed of the cusp is about one order of magnitude higher than that the relative speed of the filament systems, indicating the onset of CME is likely associated with the cusp outward motion.

\subsection{Magnetogram Images}

The above scenario of building and releasing of magnetic energy and the rearranging of field reconfiguration have also been revealed by images taken from MDI, onboard SOHO spacecraft. Sequence of images observed by MDI during and after the flare events clearly reveals the rotation of two sunspots of opposite polarities (north spot moving to the west and the south one moving to the east) at the flare site. Figure 8 shows representative MDI magnetograms in the time interval 01:30--11:15 UT. We observe a relative shift in sunspot position $\sim$4--5$\times$10$^3$ km, with an average positional shift $\sim$1000 km hr$^{-1}$. However, the positive polarity sunspot shows rapid change in the position as well as rotation during 03:00--05:00 UT (refer to Figure 10). 
In Figure 9, MDI contours over plotted on the H$\alpha$ show the typical location of the flare, which is in agreement with sunspot rotation location. Further, the X-ray location of the flare also nearly coincides on the above MDI contours.
It may be noted that the recent study has suggested that the rotation of sunspots is likely to be associated with the dynamics caused by the emerging twisted field lines (Min and Chae 2009). Thus the above magnetogram analysis reveals a considerable amount of flux emergence around the positive polarity sunspot in the time interval of 03:00--10:00 UT. However, the negative sunspot shows the rapid evolution after 08:00 UT. For comparison, we also include the shear map of the active region obtained from Marshal Space Flight Centre (MSFC) observed on 19 November 2003 at 19:36 UT (refer to Figure 9). A maximum shear along the current sheet between these opposite polarity sunspots has been recorded, and it probably caused the formation of twisted filaments.   
 The shear motions and rotations of sunspots represent the energy building process, which are likely associated with the flux emergence and cancellation (e.g., Kurokawa 1987a, Kurokawa et al. 1987b, Liu and Zhang 2001). It is consistent with destabilization of filament systems and their relative motion, leading to the formation of flares and CMEs. The magnetic cloud associated with flux rope ejection has been confirmed by interplanetary measurements (see section 2.5). Figure 11 shows the schematic cartoons to explain the second event scenario i.e.  (a) curved filaments with sunspot rotation (b) filament destabilization and interaction to produce the flare (c) cusp formation, and (d) CME eruption.

\subsection {Radio Measurements}

The above two events have also been covered well by radio observations over a range of frequency bands, 245--15400 MHz. The 1-sec radio flux density data in the above frequencies is displayed in Figure 12, individually for these two events. These plots have been made on same vertical scale for an easy comparison of strength of radio flux densities. It is evident in the figure that (i) the second event is complex and more intense than the first one. (ii) the metric and high frequency radio spectra during the flare event show moderate increase for the first event whereas huge enhancement can be seen for the second event. (iii) Moreover, the second event shows enhancement at high frequencies (8800 and 15000 MHz) on or after the flare maximum. It suggests that the much harder particle acceleration has resulted during the reconnection process of the flare event. Additionally, the dynamic spectra for the two-flare intervals are shown in Figure 13. As shown in the above plots intense type III bursts have been observed at the flare maxima, respectively, at 02:50 and 07:40 UT. The bunch of intense type III bursts during the onset of flares confirms the opening of field lines and acceleration of particles along them, resulting from the merging/reconnection caused by the approaching filament systems. In the frequency range 245--610 MHz, the radio flux density profiles show large variations over a period of about 5--10 minutes. Whereas, the radio intensity profiles in the frequency band 1415--15400 MHz, show increase over a period of about 1--2 minutes then a gradual decrease with time. The flux density peak shows systematic offset to later time with the increasing frequency.

During the maximum of the second flare, in other words at the time of effective reconnection of filament systems, sharp spiky burst has been observed at 610 and 2695 MHz around 07:39 UT and it raised about 5--7 times above the flare background radio emission at 2695 MHz. This burst interval is indicated by two vertical dotted lines in Figure 12. However, at an intermediate frequency, 1415 MHz, the increase corresponding to the above sharp burst has been nominal and it shows slightly above the flare background. It is likely that in the population of accelerated electrons, the transition from the thermal (high frequency) to non-thermal (low frequency) state may take place at this height. Nevertheless, the important point to note in the characteristics of the sharp burst at 610 and 2695 MHz is that the profile is broad at 2695 MHz, $\sim$10 sec of full width at half maximum and narrow at 610 MHz having an equivalent width of $\sim$5 sec. Moreover, the 610 MHz emission peak $\sim$3 sec later than the above frequency profile. The above results suggest that the reconnection between the filaments at lower heights (i.e. at 2695 MHz at lower corona) has caused, the production of copious amount of thermal particles over a larger volume (as the reconnection got initiated). But, the non-thermal emission (at 610 MHz) at the maximum phase of the flare has resulted from the channelization of accelerated particles along the field lines into the solar wind at height above the reconnection site. Then, the particle channel tends to cause intense but narrow width radio intensity profile. The appearance of the narrow spiky emission is consistent with H$\alpha$ profile (refer to Figure 2). The results also indicate that for the given approaching speed of the filament systems and the duration of burst, the cusp top source would size about $\sim$1000 km and the cross-section of field lines leading to non-thermal radiation is much smaller than the above.

The high frequency radio measurements, made with the Nobeyama Radio Heliograph (NoRH) are consistent with the H$\alpha$ observations and radio results described above. The NoRH could cover only the first flare event. The rising of cusp after the reconnection can be seen in both 17 and 34 GHz images as a bright source, which showed systematic movement with time. The source is brightest at $\sim$01:54 UT near the flare maximum. At 17 GHz, the foot-point sources in association with the flare ribbons could also be observed. The foot-point sources show considerable polarization, which gives a clue on the field orientation at the end of the loop. The NoRH contours are shown on H$\alpha$ and MDI magnetogram to show the relative position of the flare in different observations (refer to Figure 4).

\subsection{EUV and White-light Observations}
In association with the above flare events, two individual CMEs have been observed in the interplanetary medium. We also carefully examined the CME initiation using Extreme Ultra-Violet Telescope (EIT) images observed at 195 \AA \ and it revealed that the onset of the second event was associated with dimming above the flare site showing material depletion (e.g., Zarro et al. 1999). The  consecutive  EIT images showed movement of filaments as seen in H$\alpha$ images.

The white light images made by the Large Angle and Spectrometric Coronagraph (LASCO) C2 and C3 telescopes onboard SOHO mission, provided sequence of images of CMEs corresponding to these flare events propagating in the south-west direction. In the C2 field of view, which covers $\sim$2--6 R$_\odot$, the first CME appeared at 02:48 UT and its subsequent appearances in the C2 and C3 fields of view provided linear speed of 364 km s$^{-1}$ in the sky plane within about 15 R$_\odot$, from the centre of the Sun. This CME covered $\sim$60$^\circ$ width along the position angle $\sim$220$^\circ$. However, the second CME was faster than the above one and it became a full halo event in the LASCO field of view. Some representative images of these CMEs recorded by the LASCO are shown in Figure 14. The second CME event appeared at the C2 field of view at 08:06 UT and its linear propagation speed was 669 km s$^{-1}$. It should be noted that the second event is more intense than the first one (refer to Figure 12, section 2.3). When we compare the CME initiation from the cusp rise and its follow up in the near-Sun region within LASCO field of view, it is clearly seen that the CME has gone through a heavy acceleration, supported by the expansion of the magnetic flux rope system erupted at the time of CME onset. We also see the cusp mass eruption in H$\alpha$ in association with CME onset. However, the acceleration in H$\alpha$ has been enormous compare to low level of acceleration in LASCO field of view and the rate of expansion evidently reveals the transfer of magnetic energy at the initial stage of the CME.

\subsection{Scintillation Images and Interplanetary Data}

The above halo CME has also been traced further out in the inner heliosphere using interplanetary scintillation (IPS) observation made at the Ooty Radio Telescope, operating at 327 MHz (e.g., Manoharan et al. 2001, Manoharan 2006). Figure 15 shows the three-dimensional tomographic reconstruction of the 3-AU heliosphere obtained from the Ooty IPS measurements on November 23, 2003. The ecliptic plane images show the density and speed structures associated with the CME. However, the three-dimensional view plots for two different orientations of Earth show inclination of the CME front with respect to the ecliptic plane. That is, the flux rope is oriented $>$50$^\circ$ to the ecliptic plane. It is in agreement with the orientation of CME observed at 1 AU as well as the moderate magnetic storms produced by the CME at Earth's orbit. The speed measurements by the IPS technique at $\sim$0.5 AU is in agreement with the speed of the interplanetary CME at 1 AU. Figure 15 shows the solar wind parameter associated with the shocks and CME at the near-Earth spacecraft (http://nssdc.gsfc.nasa.gov/omniweb). During the magnetic storms the B$_\theta$ rotates upto $\sim$70$^\circ$. In this plot, the arrival of shock in front of the CME is indicated by a vertical line. The speed of the CME at the near-Sun region (LASCO measurements), in the inner heliosphere (as recorded by IPS data) and at 1 AU (by in-situ measurements) show rather very little deceleration from $\sim$670 km s$^{-1}$ to $\sim$600 km s$^{-1}$ in the entire Sun-Earth distance. It suggests that the interaction of the CME with background (or ambient) solar wind is not effective or the CME could overcome the interaction. The energy within the CME could sustain the expansion. In other word, as shown by Manoharan (2006), the ejection of flux rope above the cusp contained sufficient amount of magnetic energy, which was utilized in the interplanetary medium to overcome the aerodynamical drag imposed by the background solar wind. However, the strength of southward B$_z$ component of the field associated with CME, the orientation of the flux rope, and its impact speed could produce a moderate storm, Dst $\approx$-85 nT at the Earth.  





 \section{Discussion}

This multi-wavelength study of two homologous flare events provides evidence that the opposite rotation and displacement of opposite polarity regions play a crucial role in building up the magnetic energy required for the flare process. Sunspot rotation is the primary driver of helicity production and injection into the corona (e.g., Tian et al. 2006, van Driel-Gesztelyi et al. 2002). In a recent study, Chandra et al. (2009) estimated the spatial distribution of magnetic helicity injection at the active region under consideration and showed the existence of localized positive helicity injection in the southern part of the active region. Therefore, the newly emerging flux, which plays a major role in characterizing the motion as well as the small scale reconnection. The rotation characteristics indicate a rapidly emerging flux system (Liu \& Zhang 2001). Further, as shown by H$\alpha$ data, the destabilization of the filaments strongly correlated with the rapidly emerging magnetic flux and it leads to the destabilization of filaments as well as eruption of merged filaments. It is evidently shown by the H$\alpha$ images that the magnetic energy has been pumped above the cusp at the low chromosphere. The correlation between the separation of flare ribbons and expansion of cusp structure evidently shows the large-scale reconnection, ejection of flux rope, and acceleration of particle during the cusp eruption. The radio signatures confirm the opening of field lines after reconnection and related particle acceleration. The flux rope has been observed as the magnetic cloud in the interplanetary medium.
 
The inflow velocity is an essential factor to initiate or cause the reconnection of field lines. For example, theoretically it has been shown that inflow velocity $\sim$10 km s$^{-1}$ would lead to Petschek-type reconnection (Petschek 1964). In the present study, we obtain the approaching speed of the filaments $\sim$10 km s$^{-1}$, which would nicely satisfy the initial condition for the reconnection to take place between field lines. The present study indicates that cusp geometry is of large scale ($\sim$ 4$\times$10$^4$ km) and the reconnection point may lie well above the photosphere. That is the twisted and injected helicity leads to sigmoid type structure. The projected reconnection height as seen from the cusp, $\sim$4$\times$10$^4$ km, is consistent with the earlier observations (i.e, Sui et al. 2004). This study has also indicated the typical size of reconnection scale, $\sim$1000 km. However, the accelerated particles are channeled along the field lines in a fairly narrow cross-section. The expansion of the flux rope (i.e. magnetic cloud) has aided the CME propagation and to overcome the drag force exerted by the background solar wind. For example, the drag force is proportional to the square of the velocity difference between the solar wind and CME expansion rate. The initial and arrival speeds of the CME, respectively, at near-Sun region and at 1 AU, provide evidence that the internal energy (magnetic energy) has supported to overcome the drag force (Manoharan 2006). The cusp shape suggests the formation of twisted loop as well as magnetic null at the high corona (e.g., Manoharan and Kundu 2003). The typical ejected mass amounts to $\sim$ $10^{15}$ gm, which accounts only a small fraction of mass of filament systems.  The follow-up of the CME in the Sun-Earth distance using the IPS technique shows that the flux rope is oriented $>$50$^\circ$ with respect to the ecliptic plane. Thus, although the CME originated close to the disk centre, it could only produce a moderate storm of Dst $\sim$ -85 nT at the Earth$^\prime$s magnetosphere. However, IPS and in-situ data record a strong shock associated with the CME propagation.

In summary, this study shows evidences that the sunspot motion characteristics correspond to rapid emergence of magnetic flux. The destabilization of filaments strongly correlates with sunspot motion or rotation. The eruption of the twisted-filament system suggests that the small-scale reconnection or tether cutting and the building of magnetic energy lead to the cusp formation. The inflow speed is consistent with the fast reconnection. However, the geometry and size of the cusp indicates a considerably large magnetic energy associated with the system, and it is likely provided by the helicity injection caused by the rotation of sunspots. The observed magnetic cloud confirms the associated flux rope. This study has provided a unique example to understand the space-weather consequences of solar activities in the Sun-Earth distance.   





\acknowledgments

It is a great pleasure to thank Prof. Ram Sagar for his support and encouragement in the present study. We acknowledge the
technical staff of ARIES Solar Tower Telescope for their assistance in maintaining and making regular solar observations. The authors thank the observing and engineering staff of Radio Astronomy Centre in making the IPS observations. We also thank B. Jackson and the UCSD team for the IPS tomography analysis package. SOHO (EIT, LASCO, and MDI images) is a project of international cooperation between ESA and NASA. PKM acknowledges the partial support for this study by CAWSES-India Program, which is sponsored by ISRO. We acknowledge the team of Learmonth observatory for making radio dynamic spectra available at http://www.ngdc.noaa.gov/stp/SOLAR/ftpsolarradio.html. We also acknowledge the Nobeyama team for providing the radio data used in this study.

\begin{figure*}
\centerline{\hbox{\hspace{0.1in}
\includegraphics[width=10cm]{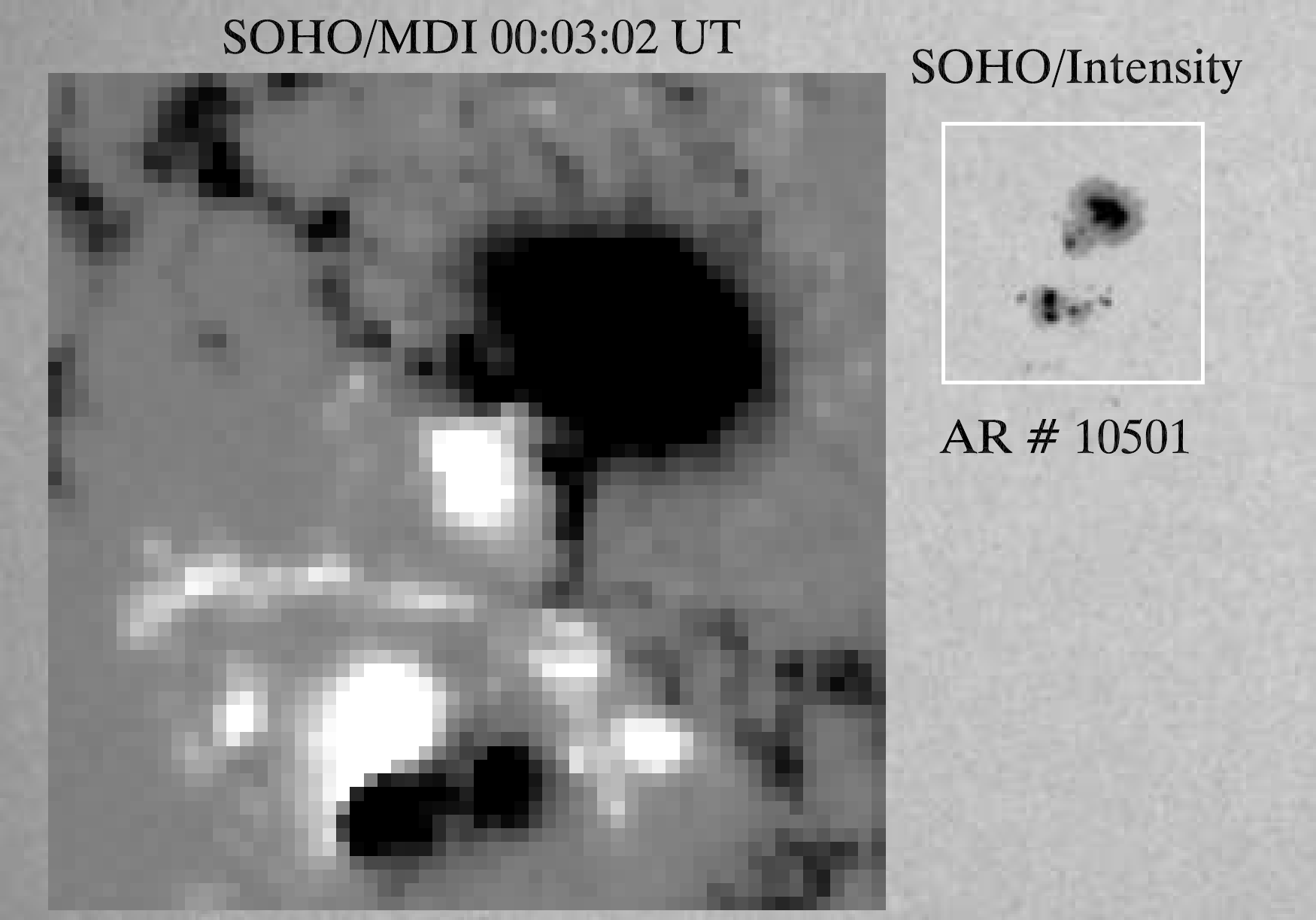}
}}
\caption{SOHO/MDI magnetogram of the active region NOAA 10501 on 20 November 2003. White light image of the active region is shown inside the box.} 
\label{fig1}
\end{figure*}

\begin{figure*}
\epsscale{1.5}
\plotone{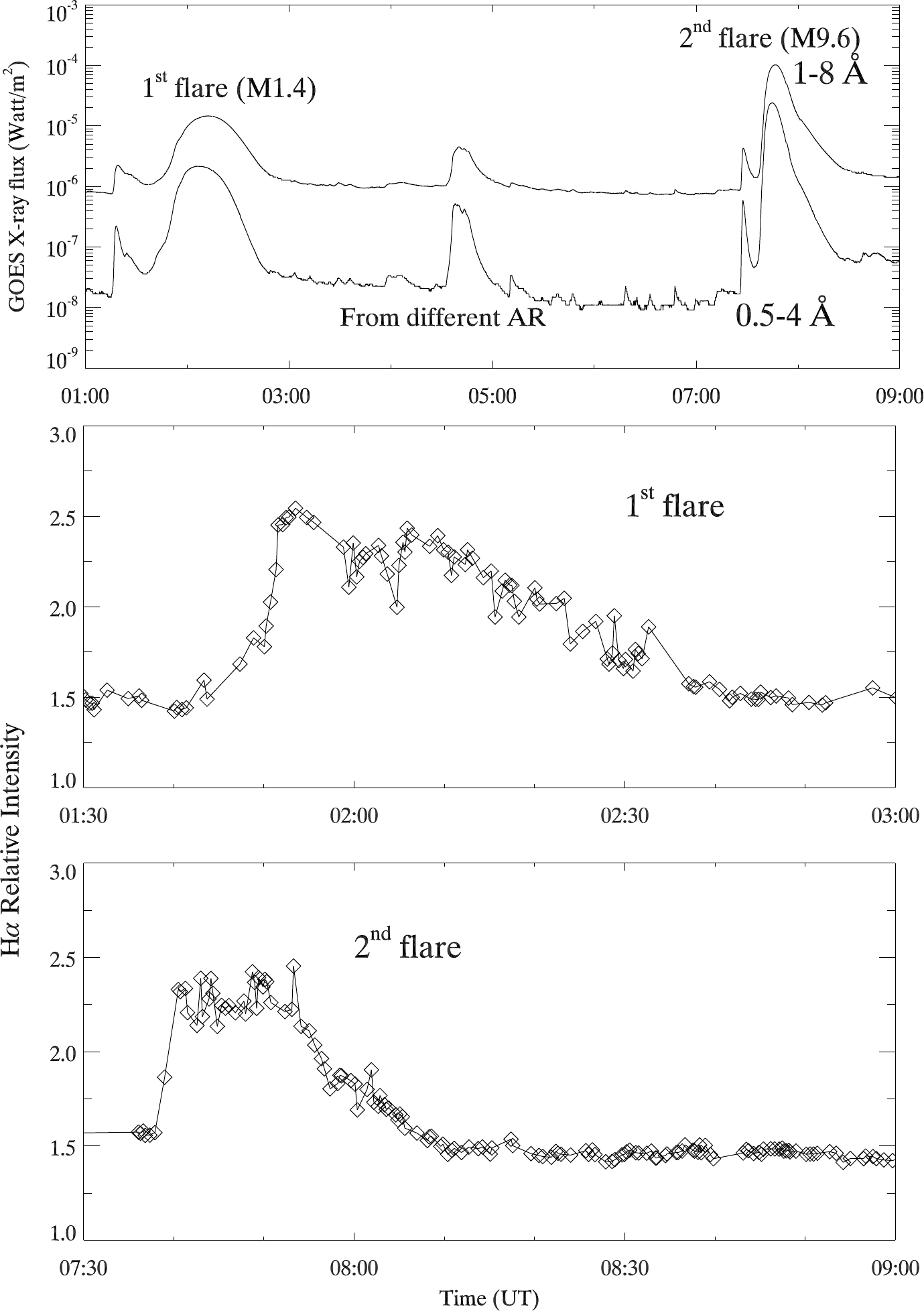}
\caption{GOES soft X-ray flux measurements in 0.5-4 \AA \  and 1-8 \AA \  wavelength bands (top) and time profiles of the H$\alpha$ relative intensity with respect to the background emission for both flares.
\label{fig2}}
\end{figure*}
\begin{figure*}
\epsscale{2.0}
\plotone{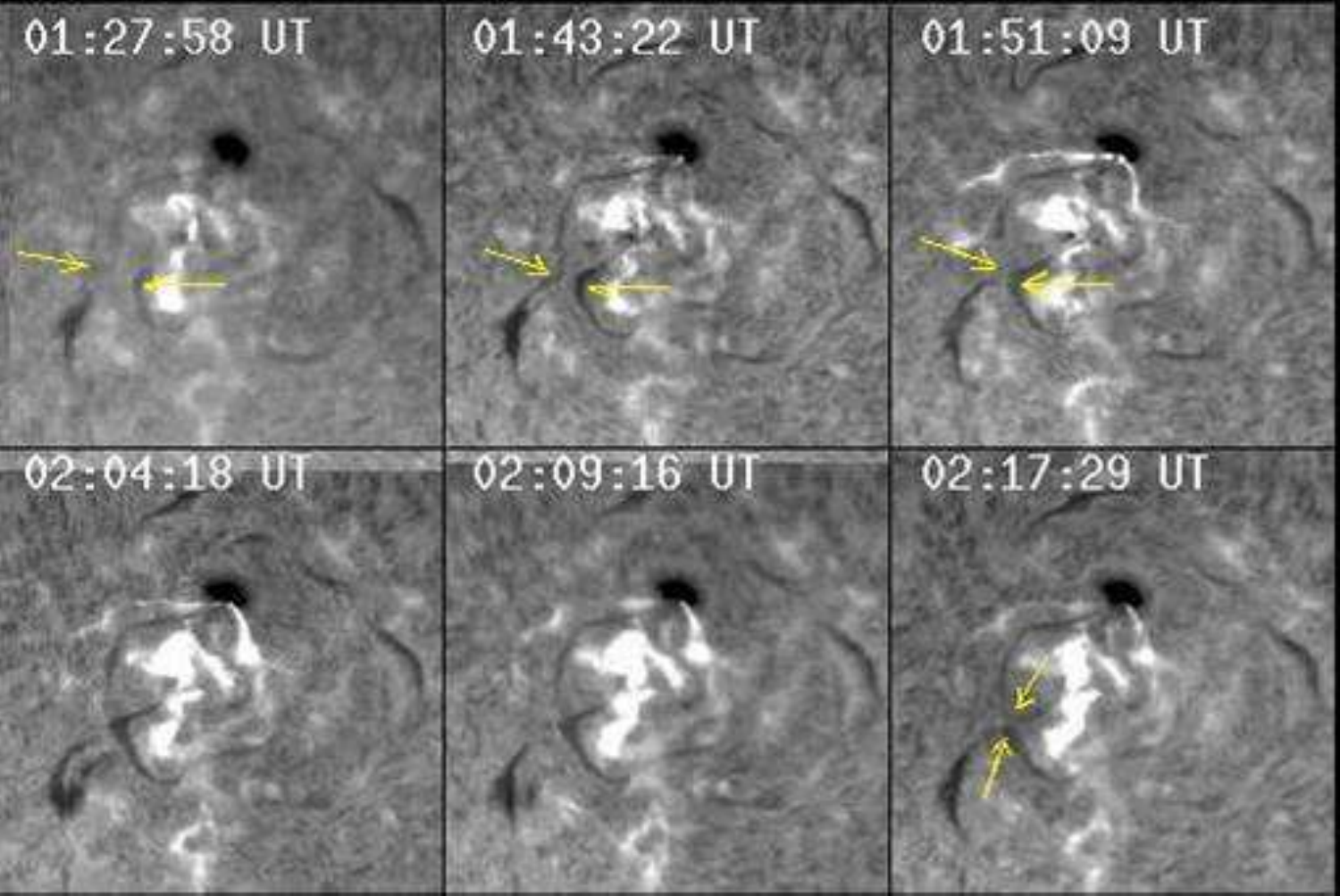}
\caption{H$\alpha$ images of the first flare (1N/M1.4) showing the evolution of filaments and their interaction. The size of each image is  315\arcsec$\times$315\arcsec.
\label{fig3}}
\end{figure*}


\begin{figure*}
\centerline{\hbox{\hspace{0.5in}
\includegraphics[width=6cm]{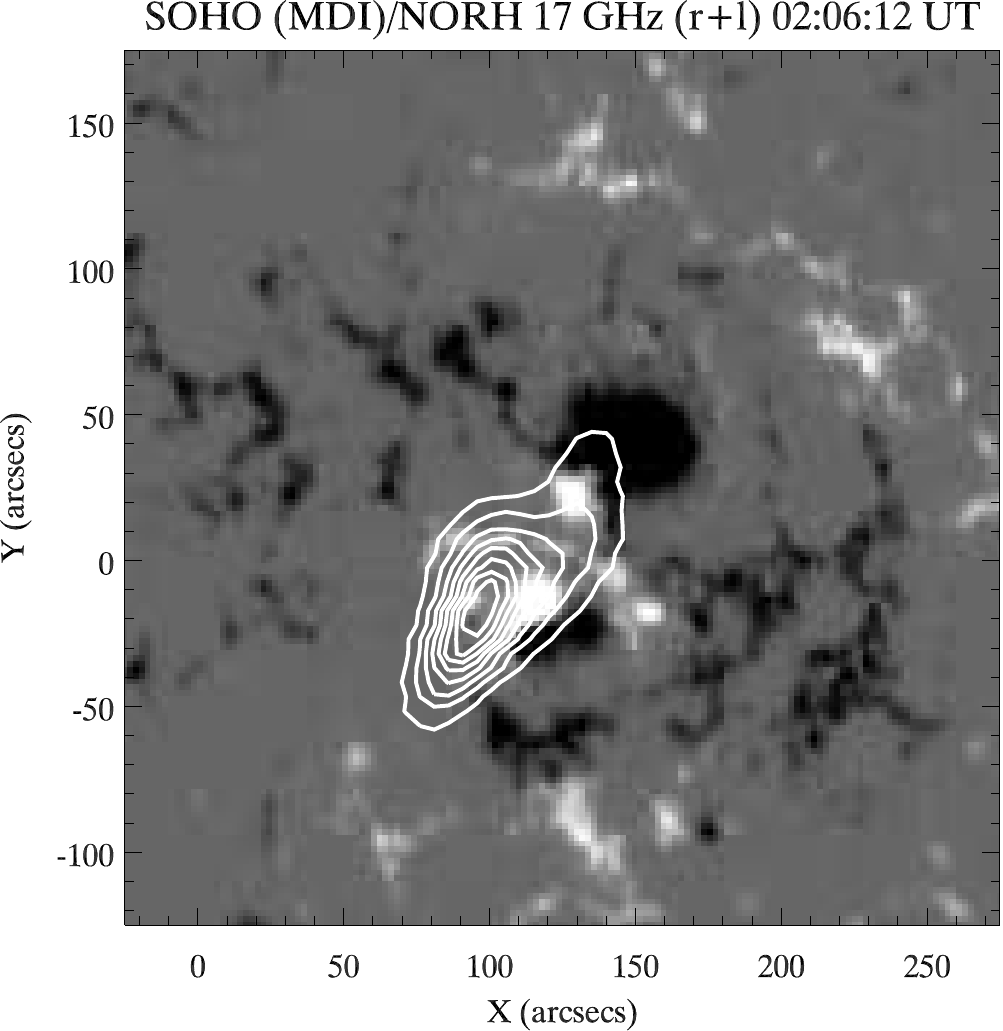}
\includegraphics[width=6cm]{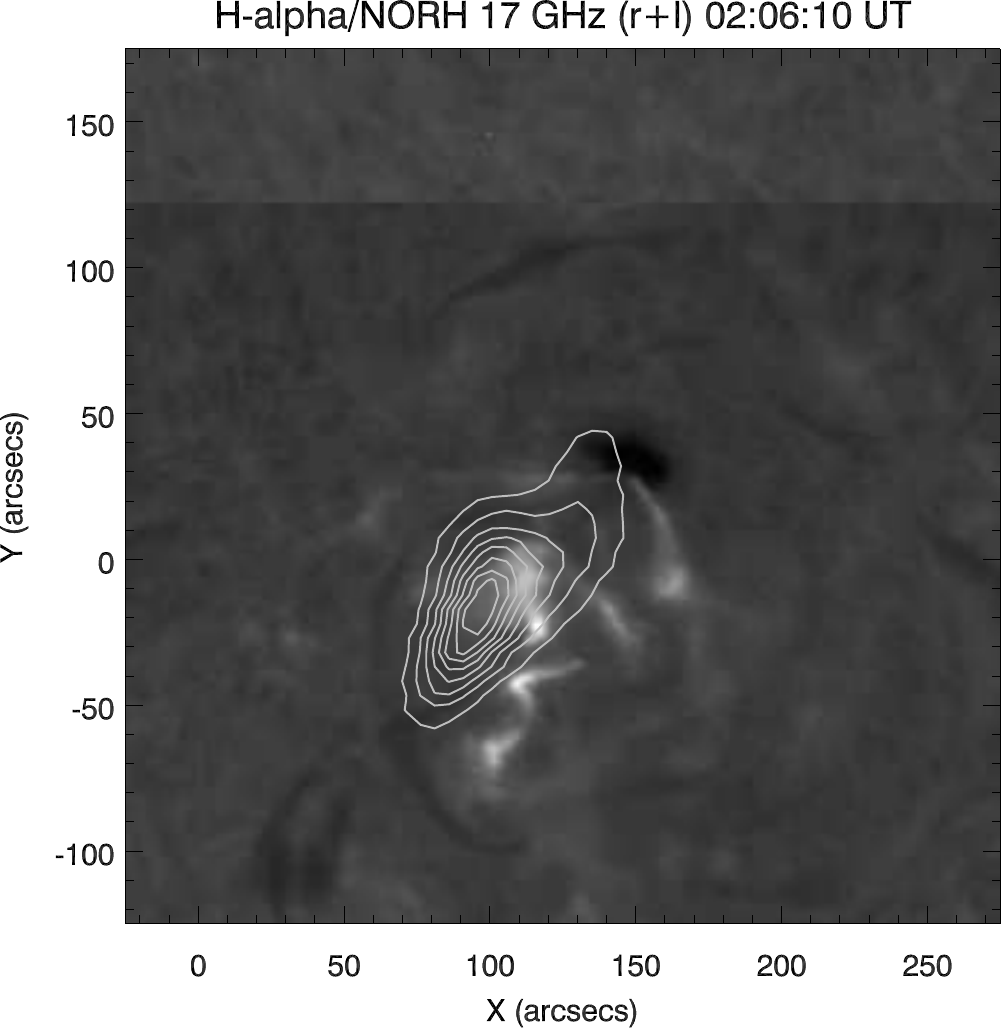}
\includegraphics[width=6cm]{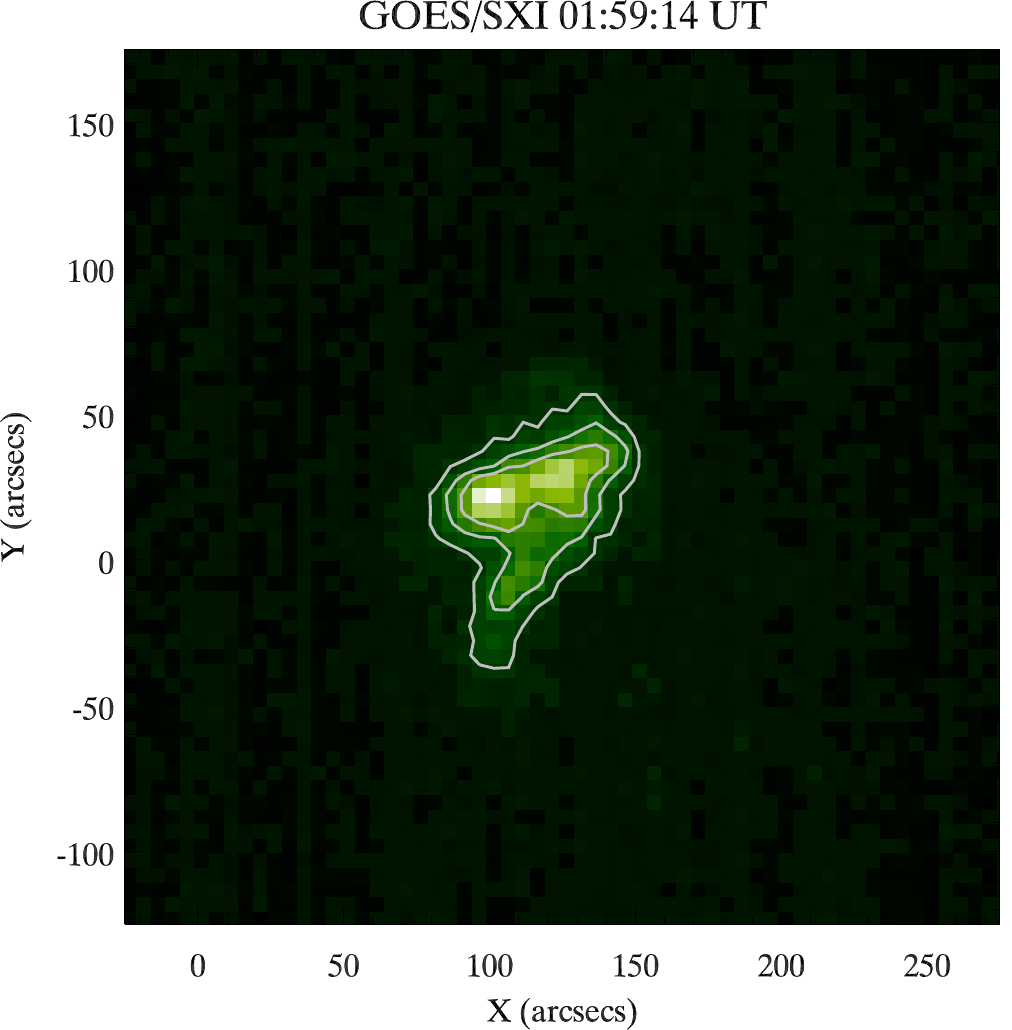}
}}
\centerline{\hbox{\hspace{0.5in}
\includegraphics[width=18cm]{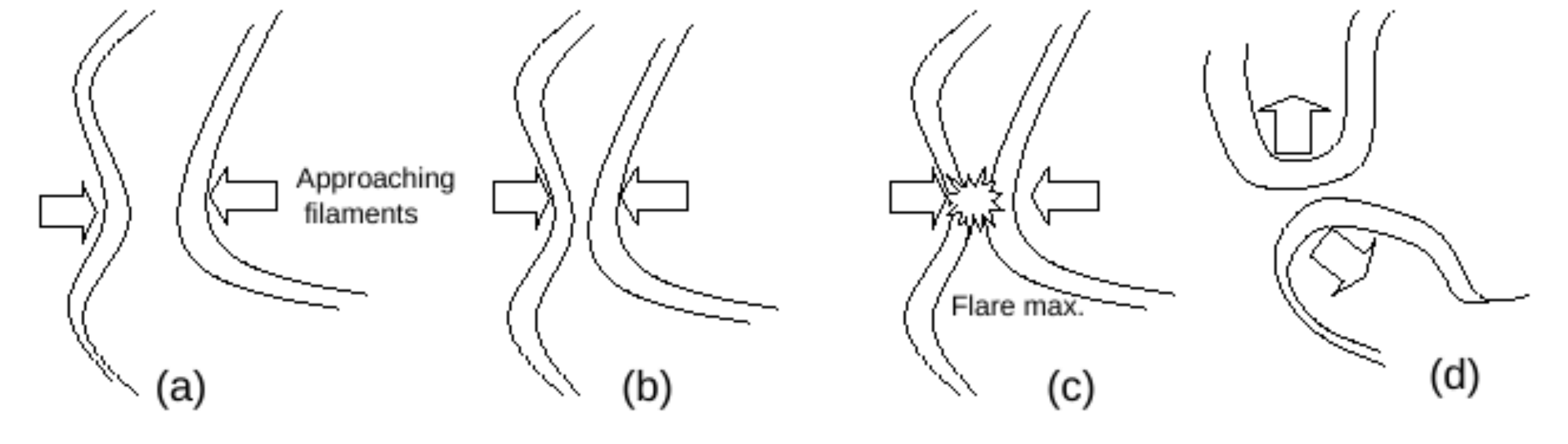}
}}
\caption{Nobeyama 17-GHz contours overlaid on MDI image (top, left) and on H$\alpha$ image (top, middle), and GOES/SXI image showing the X-ray source location (top, right). The schematic cartoons show the evolution of the first flare, i.e., approaching and interacting filaments followed by magnetic reconnection.
\label{fig4}}
\end{figure*}
\begin{figure*}
\epsscale{1.5}
\plotone{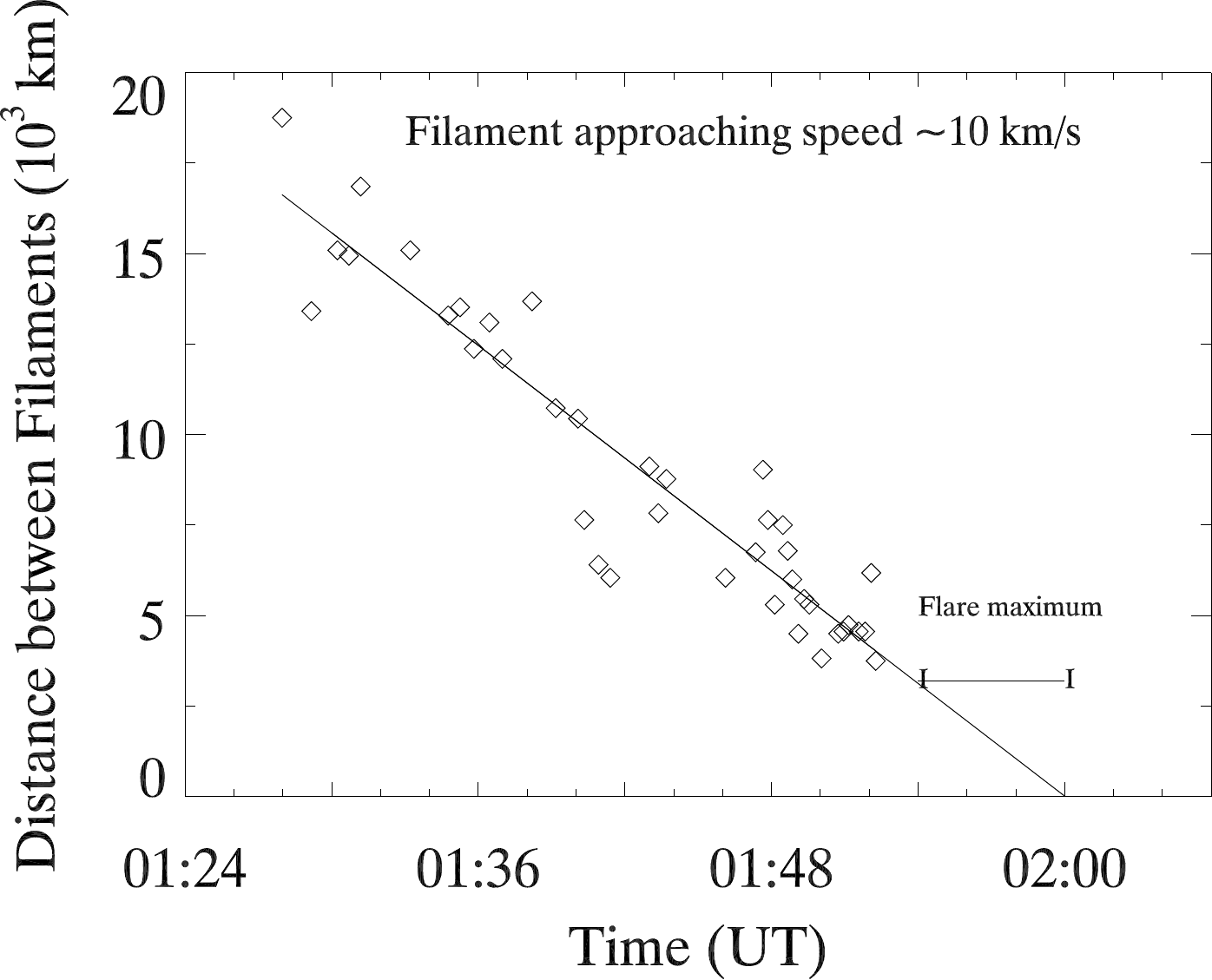}
\caption{The distance between two filaments plotted as a function of time. The straight line is the least square fit to the data points. The typical inflow speed is $\sim$10 km s$^{-1}$. The H$\alpha$ intensity of this event attains maximum between 01:53 and 02:00 UT.
\label{fig5}}
\end{figure*}
\begin{figure*}
\epsscale{2.0}
\plotone{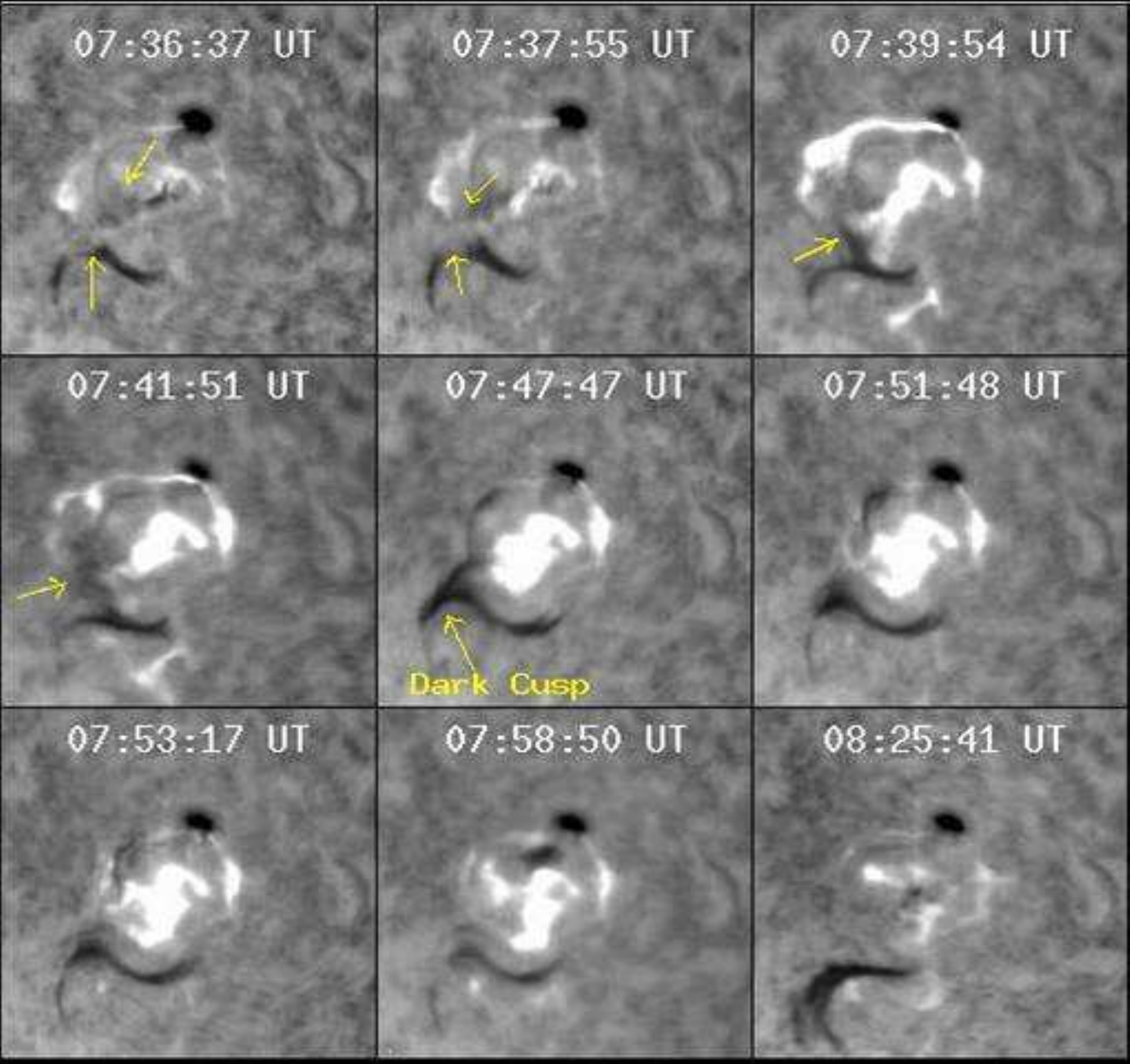}
\caption{H$\alpha$ images of the second flare (2B/M9.6) show the evolution of field lines. The dark cusp shows the mass motion at the height in the low corona. The size of each image is 315\arcsec$\times$315\arcsec.
\label{fig6}}
\end{figure*}
\begin{figure*}
\epsscale{1.5}
\plotone{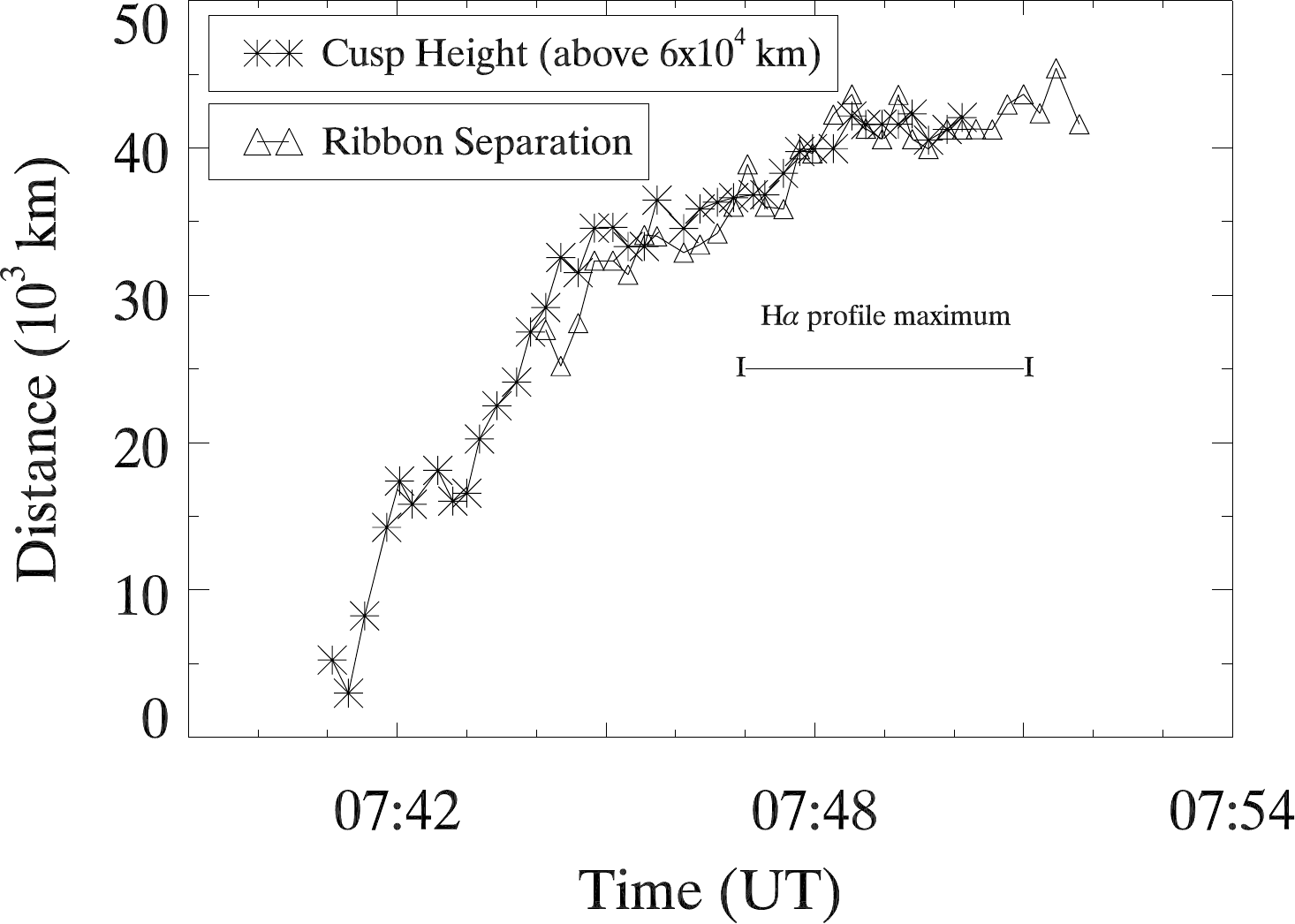}
\caption{Plot showing the cusp height variation and ribbon separation as a function of time during the flare. After $\sim$07:45 UT, the H$\alpha$ profile attains the maximum at which the cusp height stablizes.
\label{fig7}}
\end{figure*}
\begin{figure*}
\centerline{\hbox{\hspace{0.23in}
\includegraphics[width=2.0in]{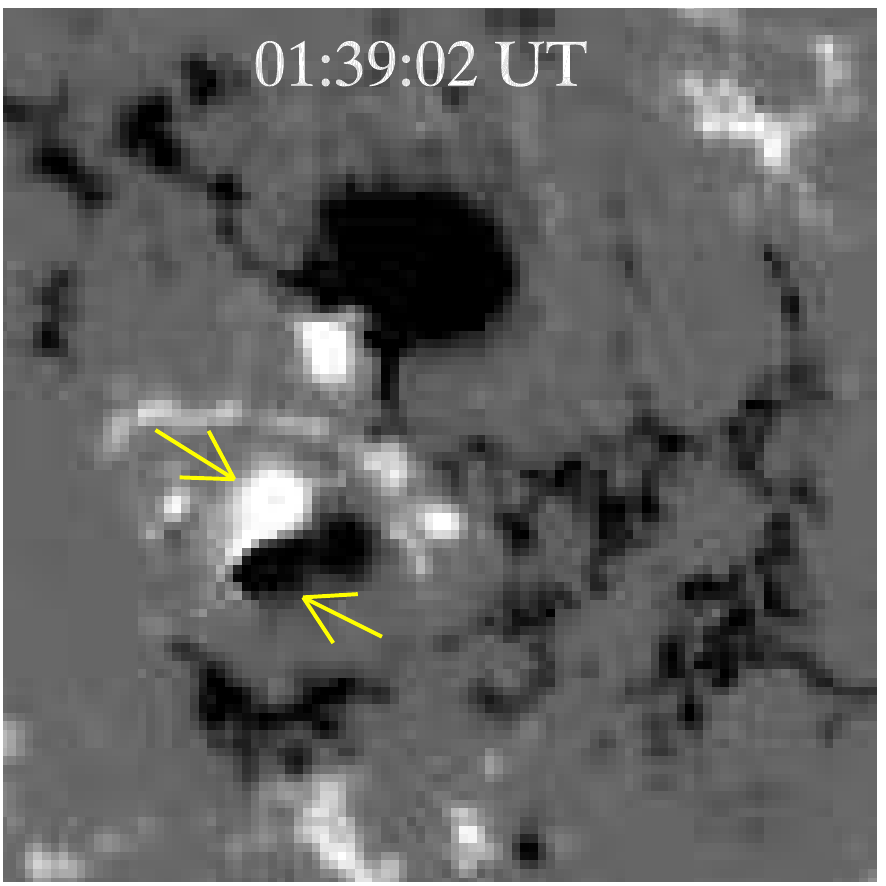}
\includegraphics[width=2.0in]{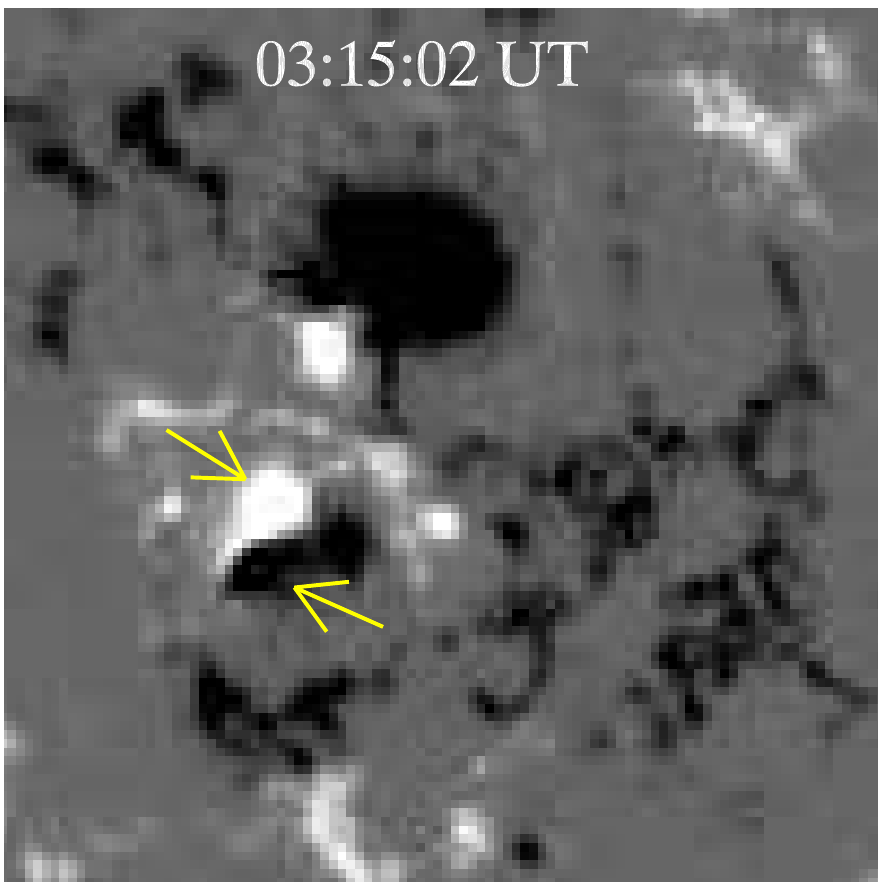}
\includegraphics[width=2.0in]{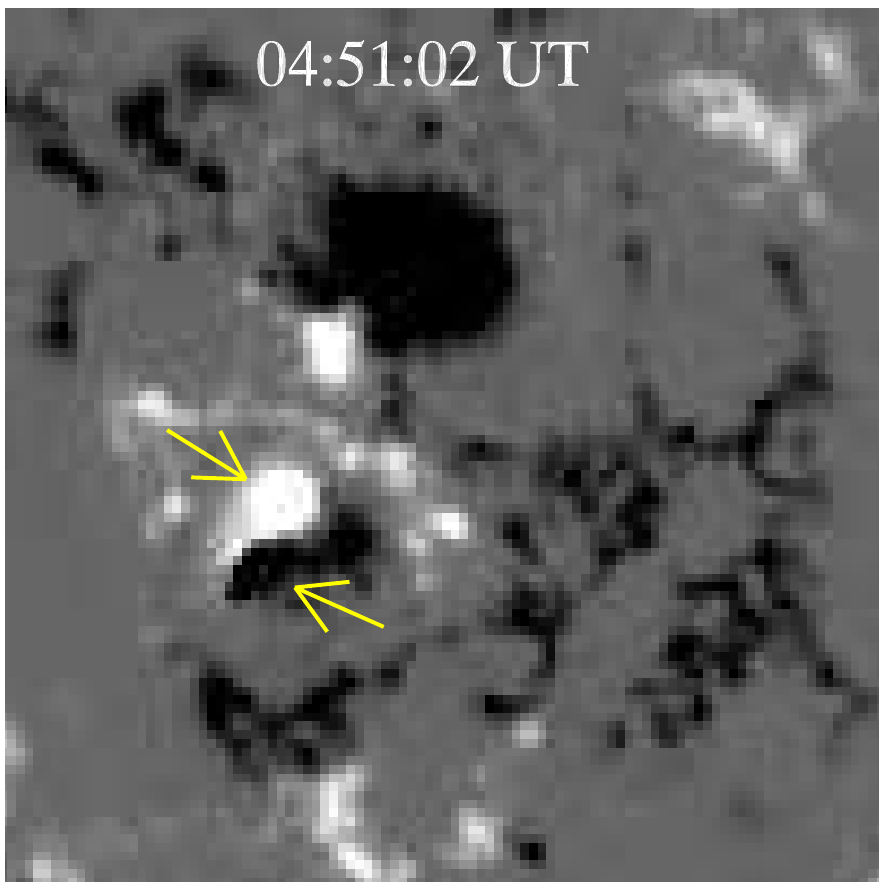}
}}
\centerline{\hbox{\hspace{0.001in}
\includegraphics[width=2.26in]{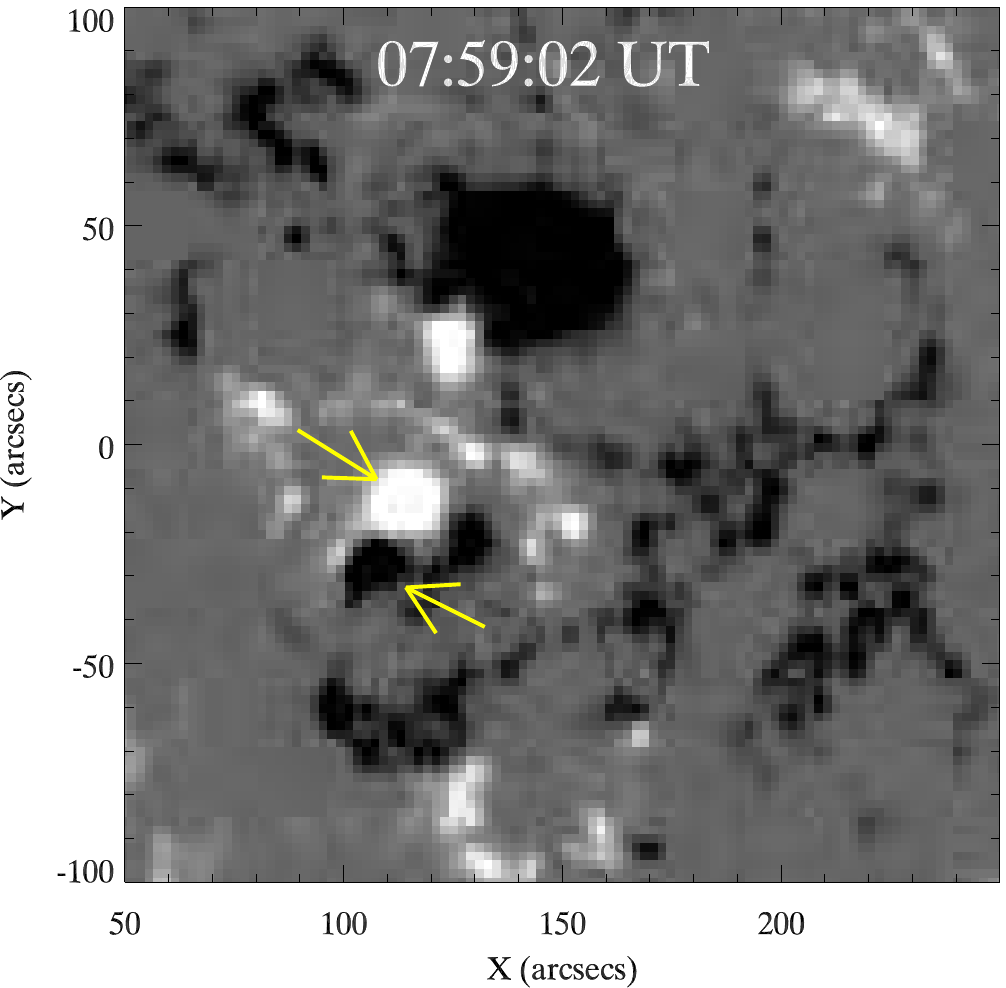}
\includegraphics[width=2.0in]{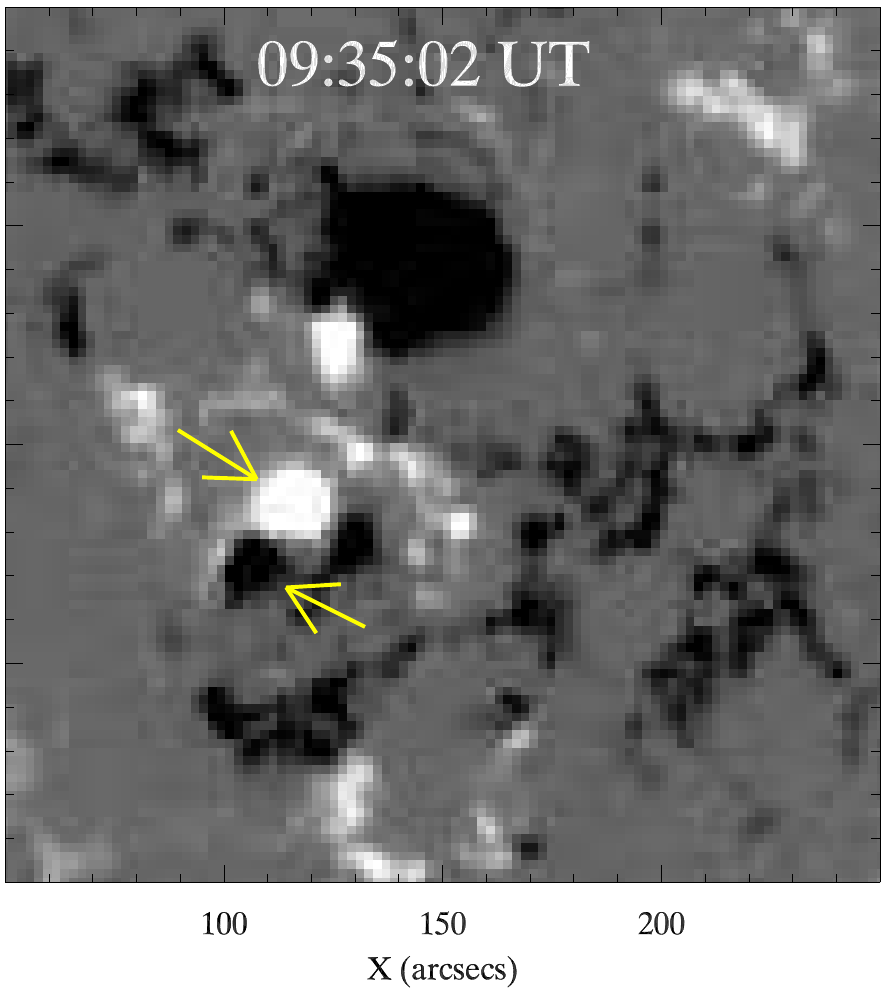}
\includegraphics[width=2.04in]{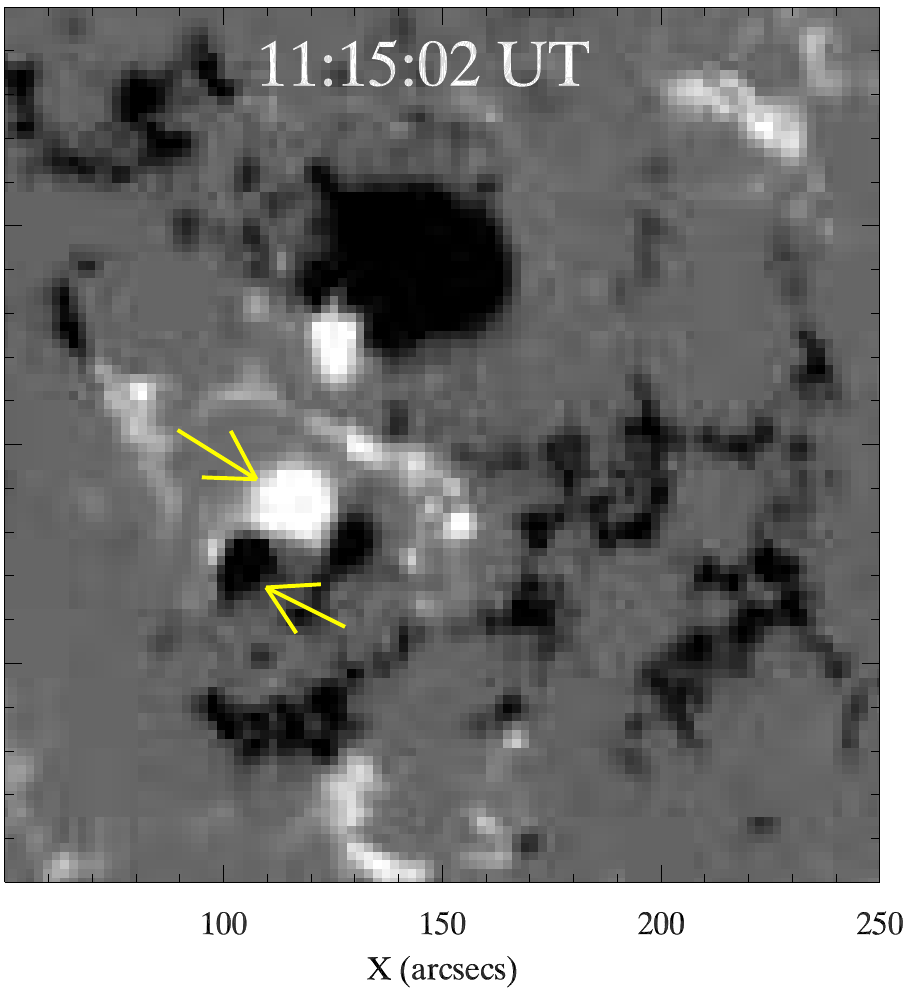}
}}
\caption{MDI magnetograms of the active region on 20 November 2003, showing the clockwise rotation of positive polarity sunspot and anticlockwise rotation of negative polarity sunspot (shown by arrows).
\label{fig8}}
\end{figure*}

\begin{figure*}
\centerline{\hbox{\hspace{0.5in}
\includegraphics[width=9cm]{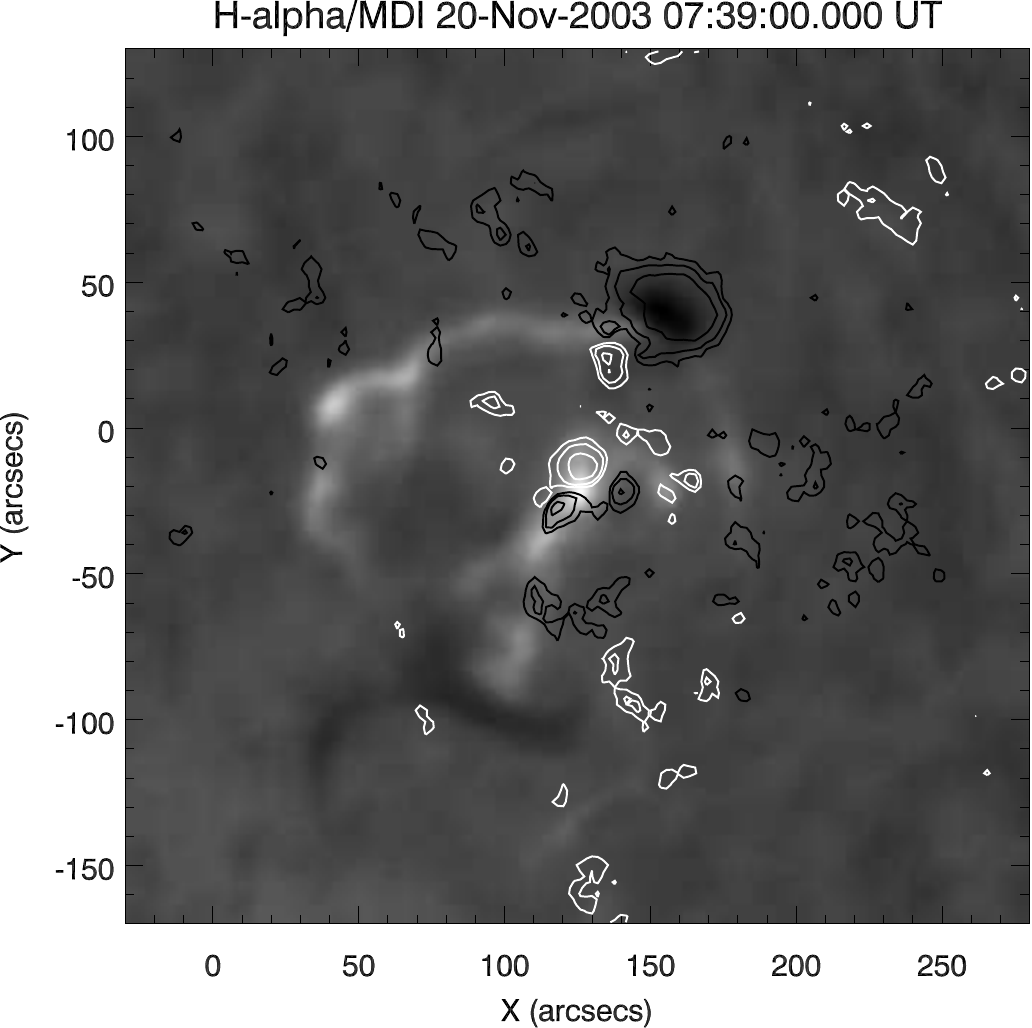}
\includegraphics[width=11cm]{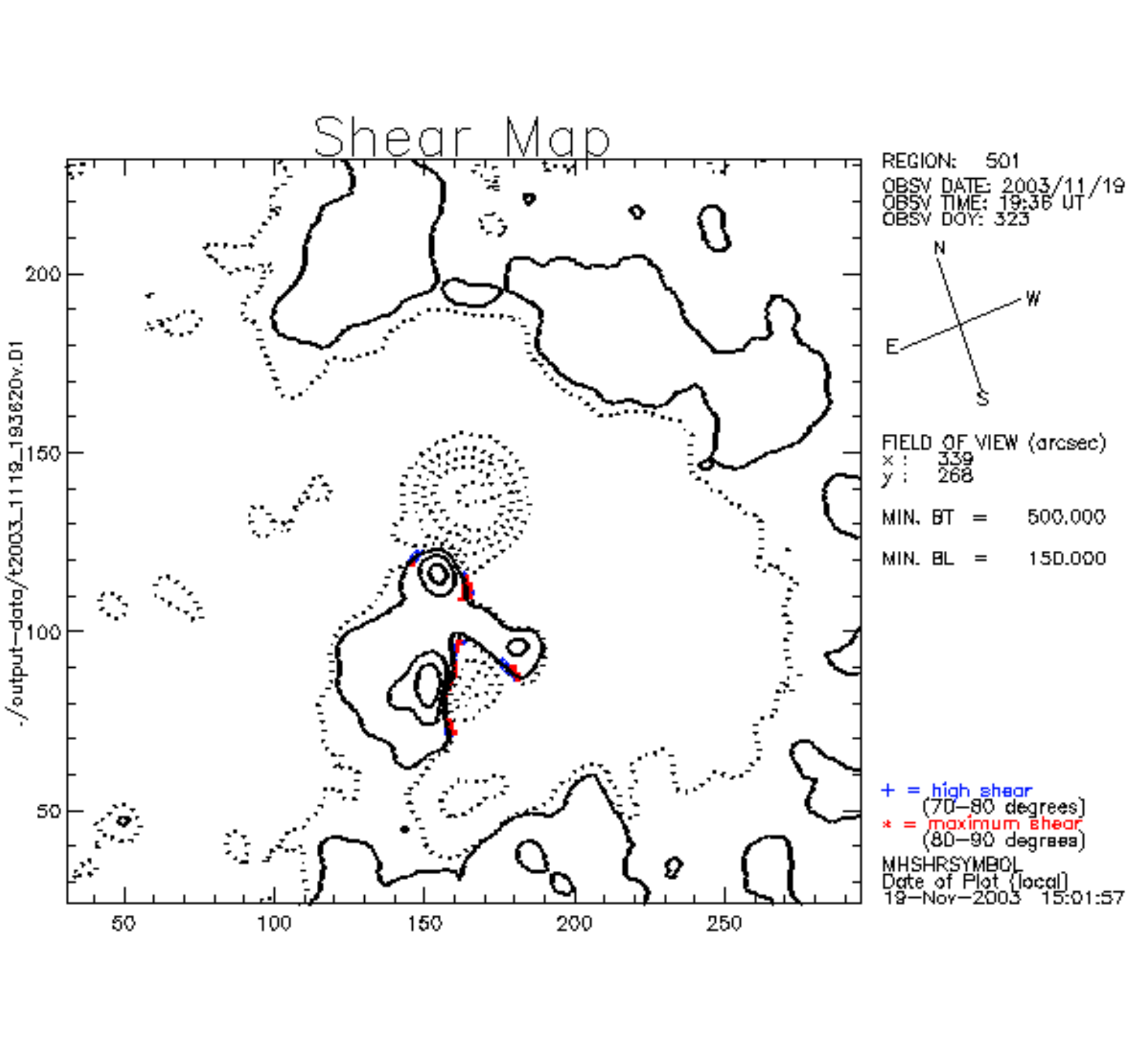}
}}
\caption{Left: MDI contours overlaid on H$\alpha$ image during the second flare event. Right: MSFC shear map of the active region showing the maximum shear at the flare site in between the sunspots on 19 November 2003 at 19:36 UT, is shown for comparison.
\label{fig9}}
\end{figure*}
\begin{figure*}
\epsscale{1.5}
\plotone{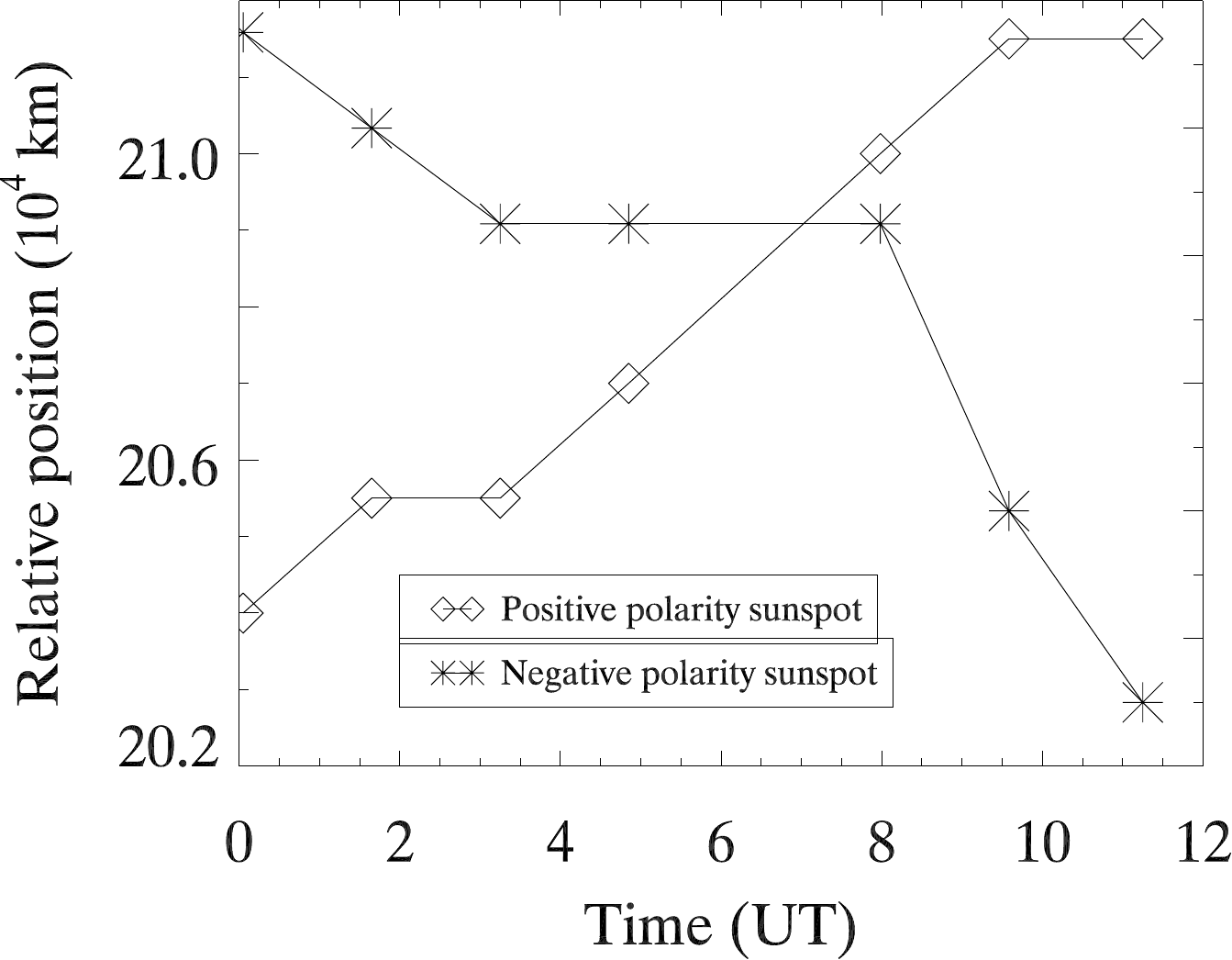}
\caption{The relative position change with time of opposite polarity sunspots showing the motion of the both sunspots.
\label{fig10}}
\end{figure*}
\begin{figure*}
\epsscale{2.0}
\plotone{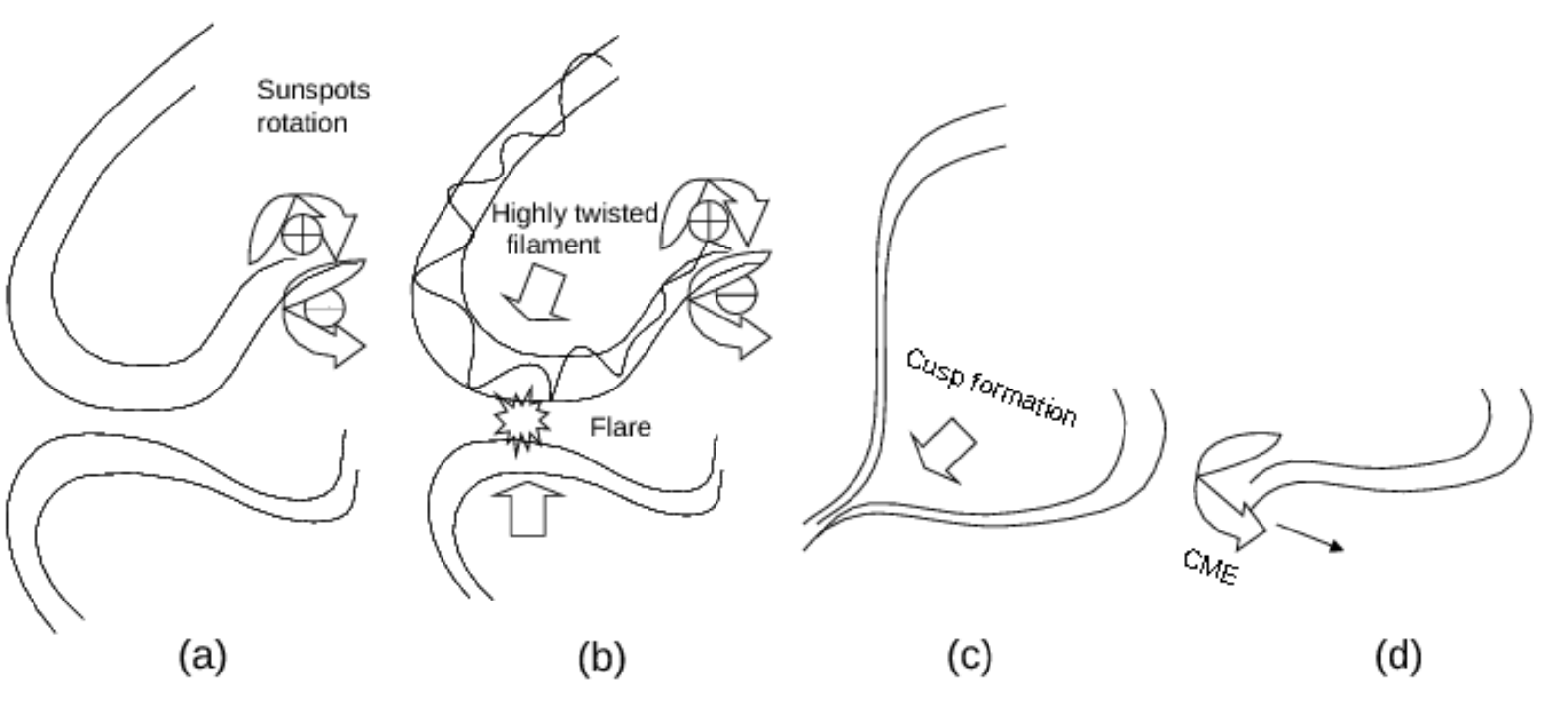}
\caption{The schematic cartoon showing the second flare evolution with one highly twisted filament destabilization in association with rotating sunspots and merging with another curved filaments,  forming a cusp and resulting CME eruption.
\label{fig11}}
\end{figure*}

\begin{figure*}
\includegraphics[width=8.5cm]{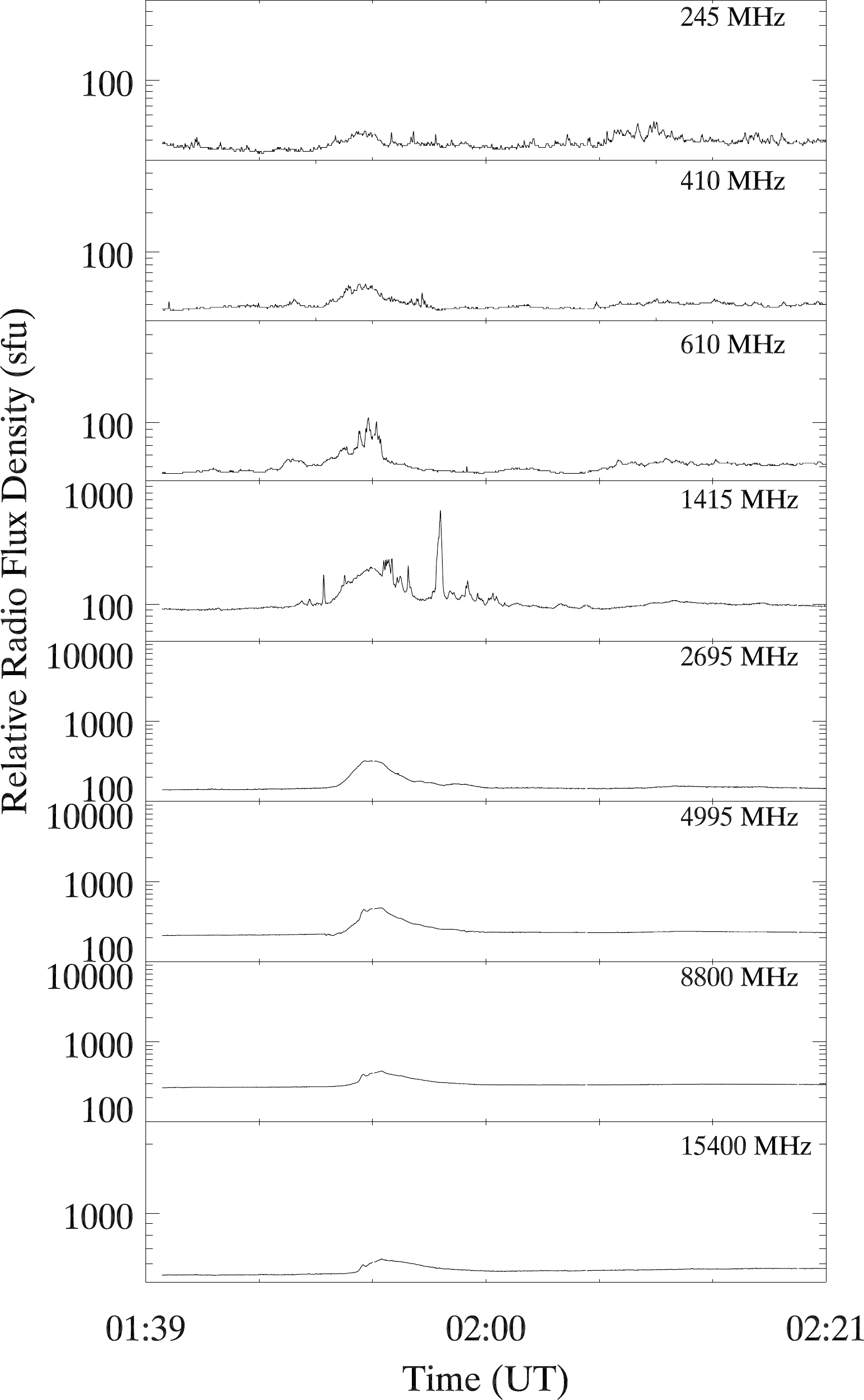}
\includegraphics[width=7.1cm]{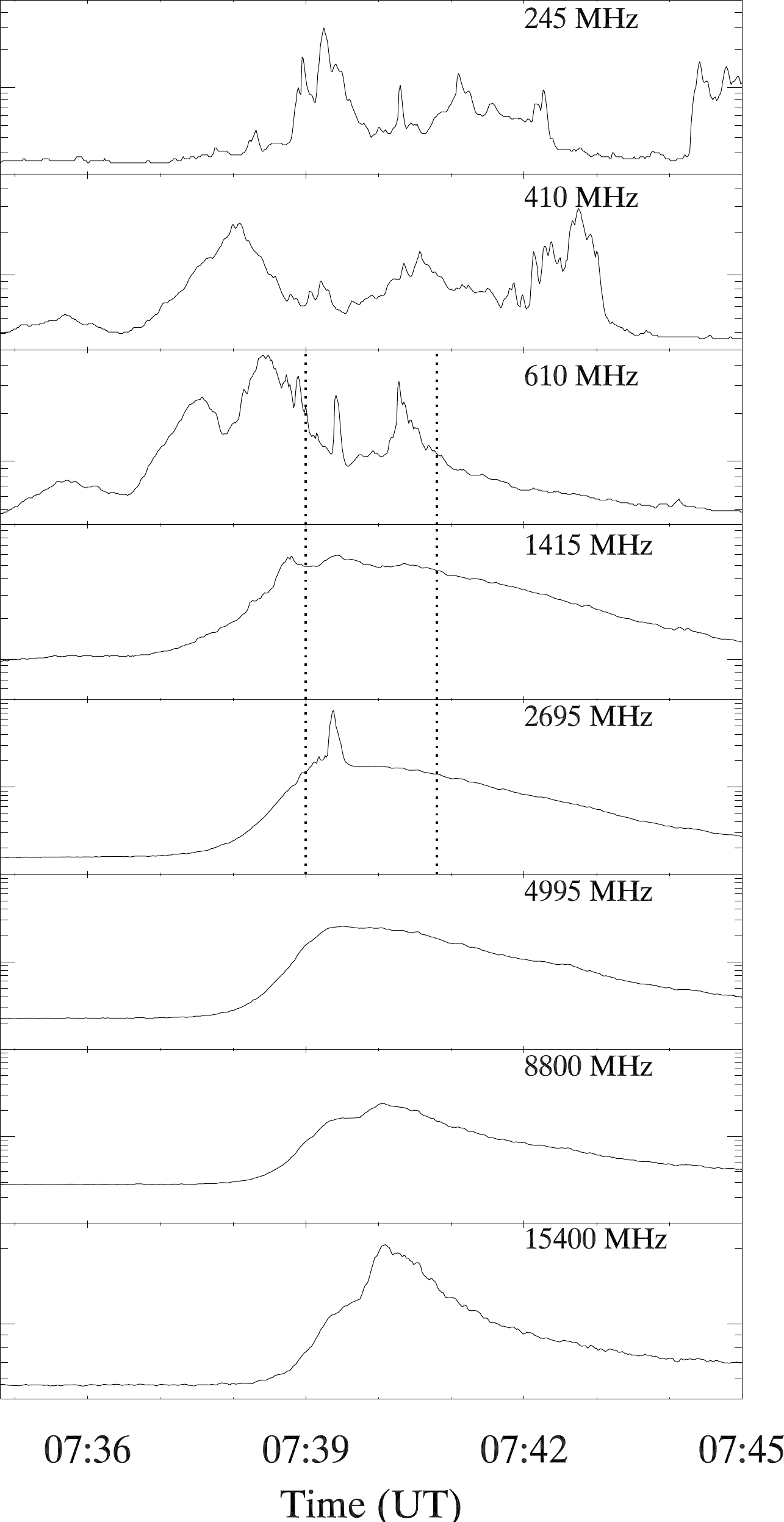}

\caption{Time profiles of radio flux densities observed at eight frequency bands from Learmonth observatory for the first and second flares, respectively.  These plots have been made in same scales for an easy comparison of strength of these flares. It is evident that second event is complex and more intense. Time profiles of radio flux density at the reconnection time (or interaction between filaments) observed at 610, 1415, and 2695 MHz is shown in between two vertical dotted lines. The time lag of $\sim$2.5 sec is evidently shown between 2695 and 610 MHz. The coronal height increases from bottom to top.
\label{fig12}}
\end{figure*}
\begin{figure*}
\centerline{\hbox{\hspace{0.5in}
\includegraphics[width=8.25cm]{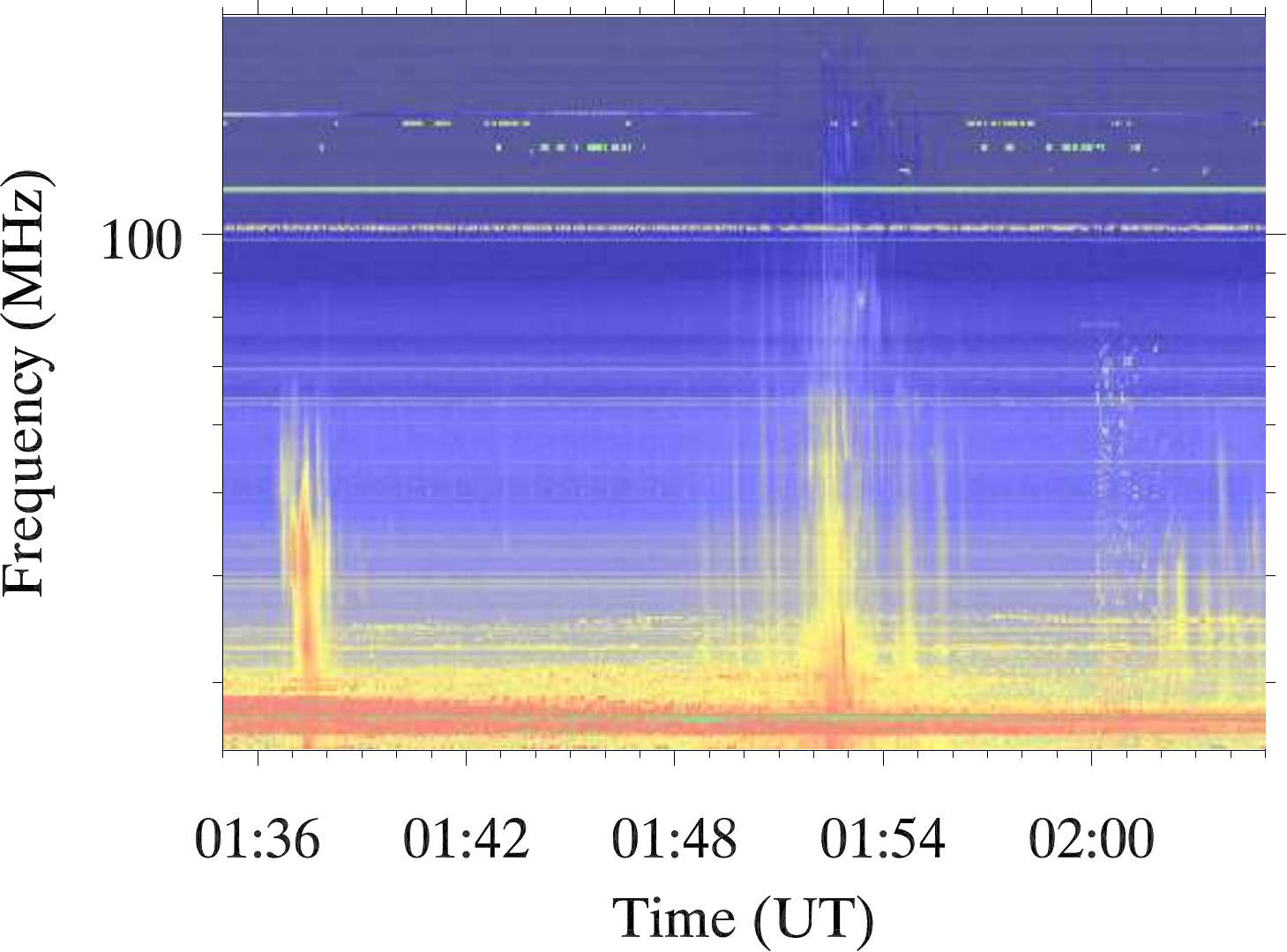}
\includegraphics[width=7cm]{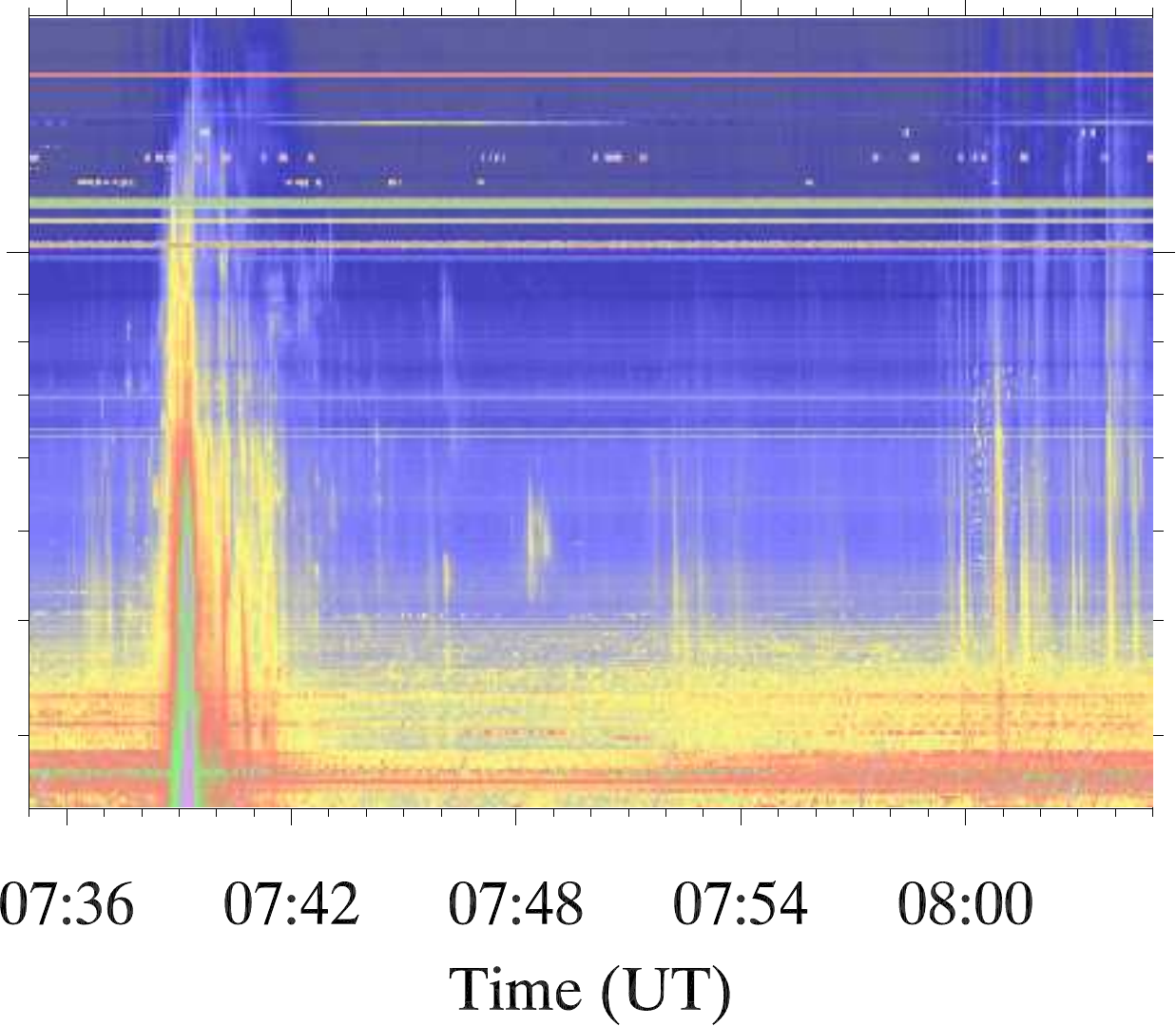}
}}
\caption{Type III bursts associated with both flare events on 20 November, 2003 (Learmonth, Australia). These intense bursts provide evidence for the opening of field lines at the time of reconnection.
\label{fig13}}
\end{figure*}

\begin{figure*}
\centerline{\hbox{\hspace{0.5in}
\includegraphics[width=2.3in]{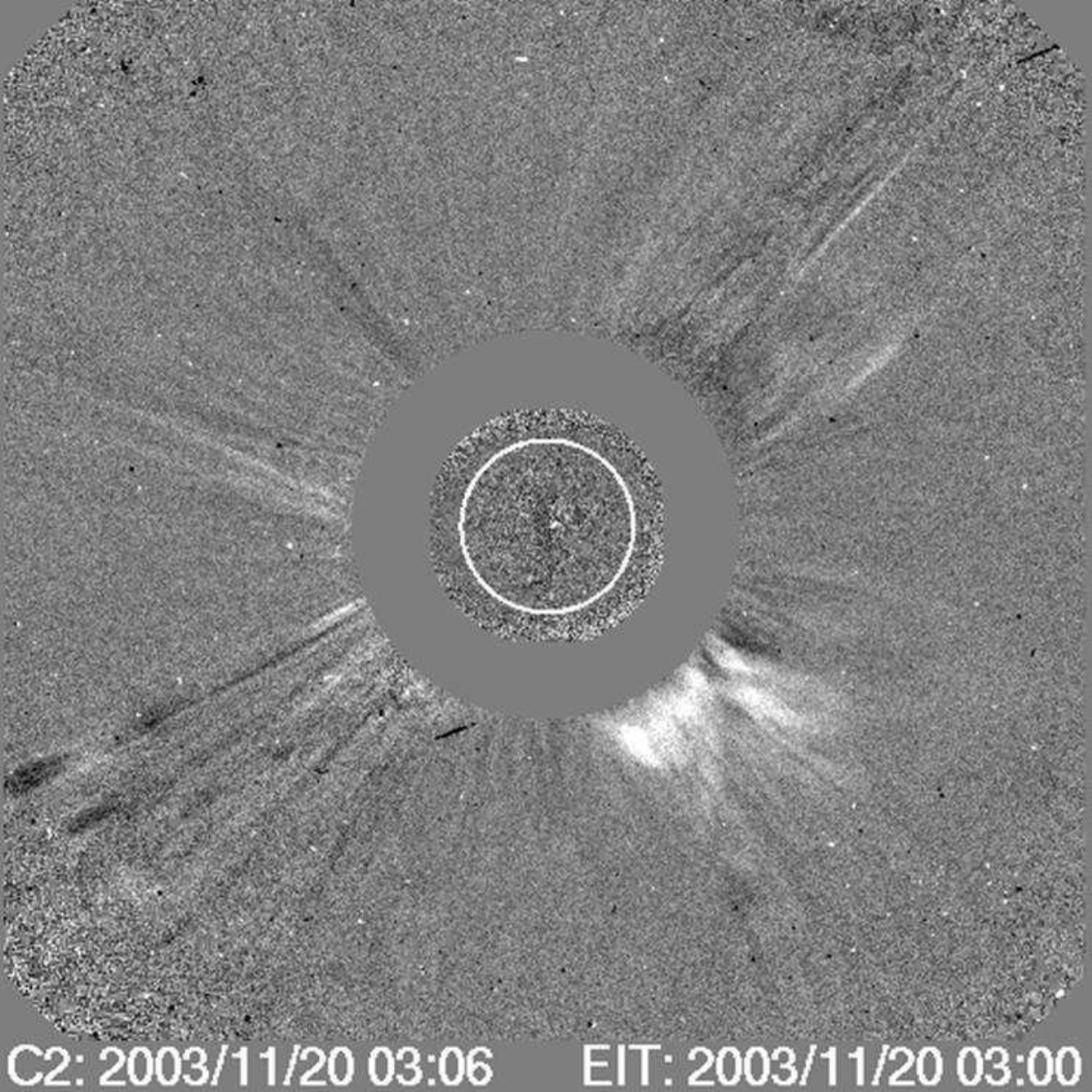}
\includegraphics[width=2.3in]{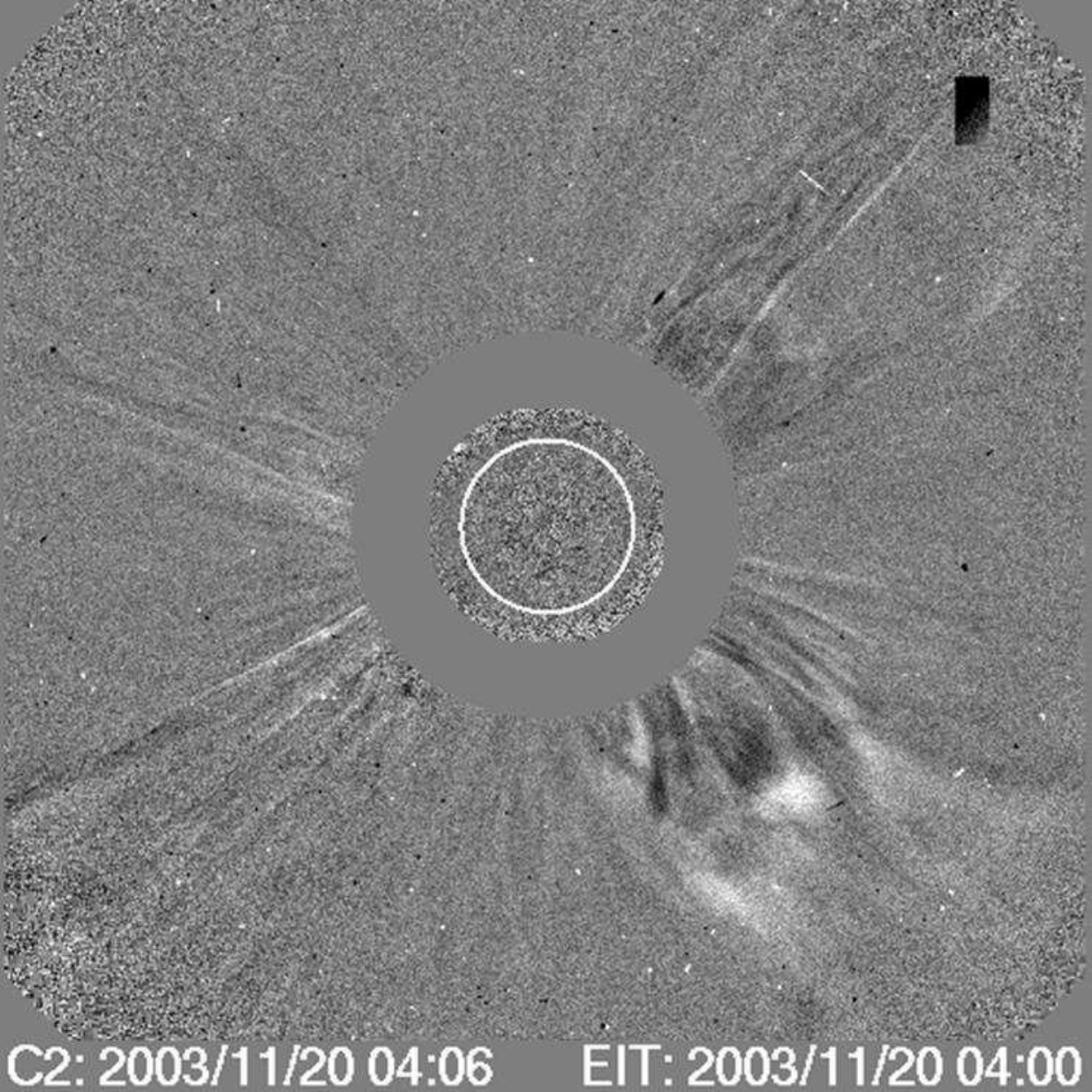}
\includegraphics[width=2.3in]{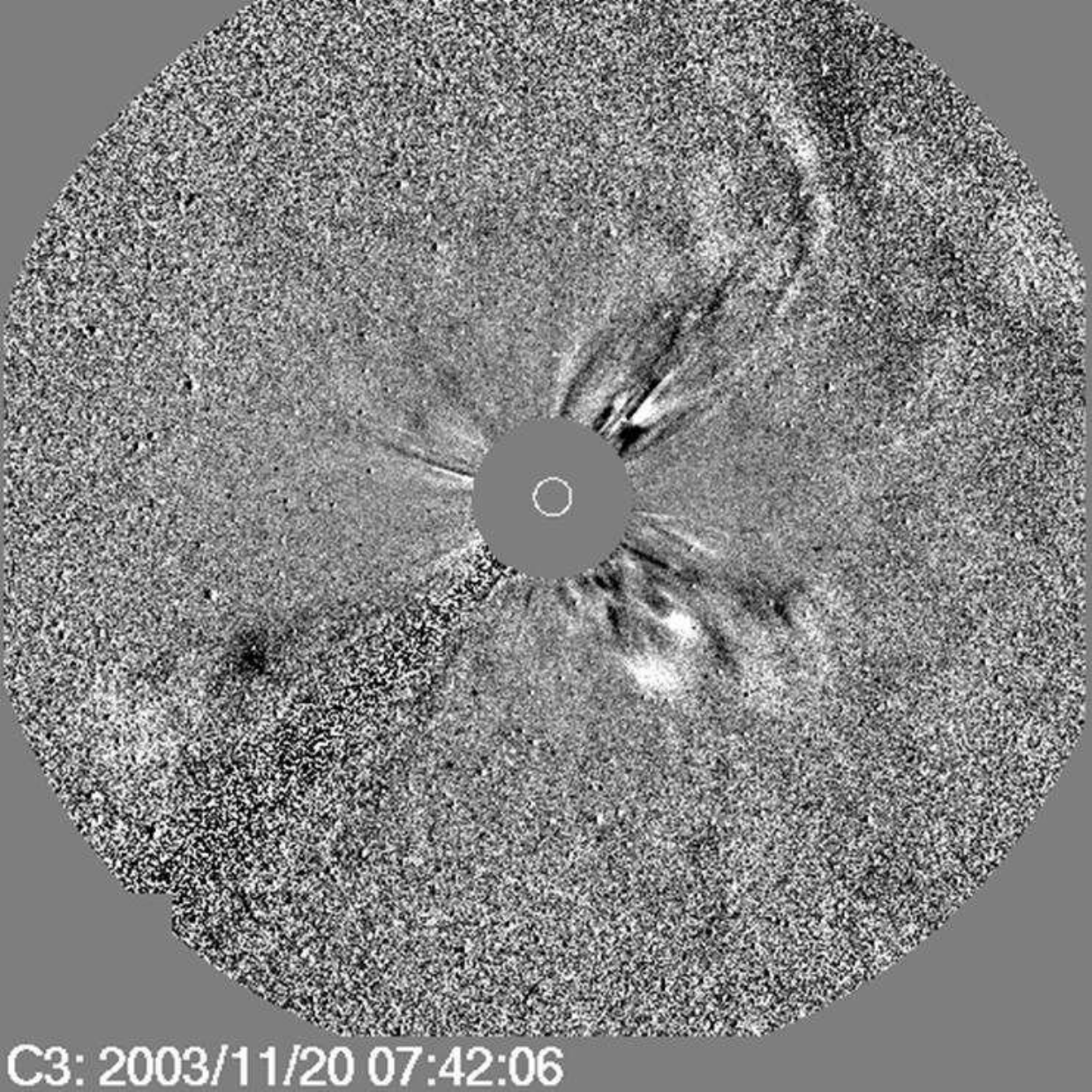}
}}
\centerline{\hbox{\hspace{0.5in}
\includegraphics[width=2.3in]{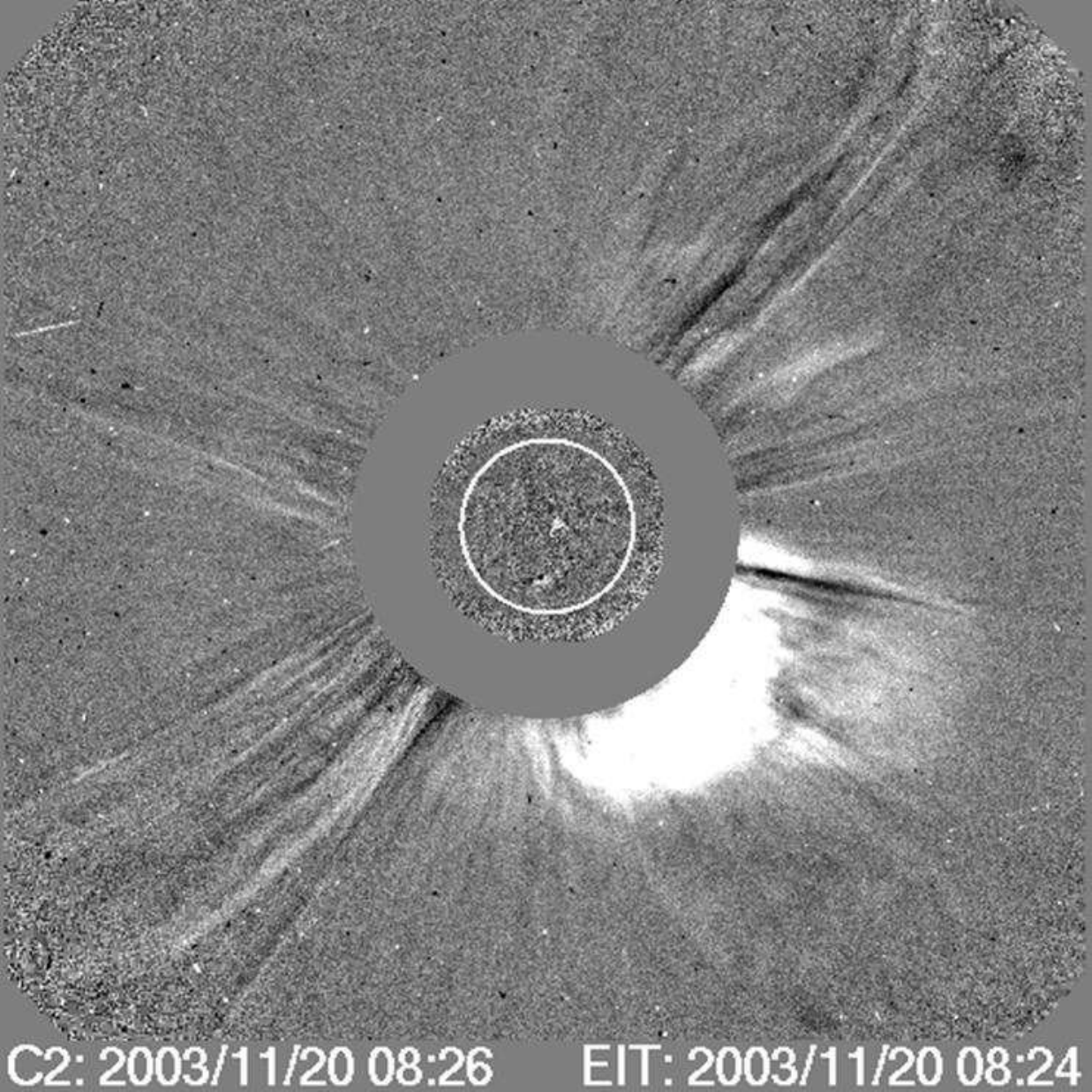}
\includegraphics[width=2.3in]{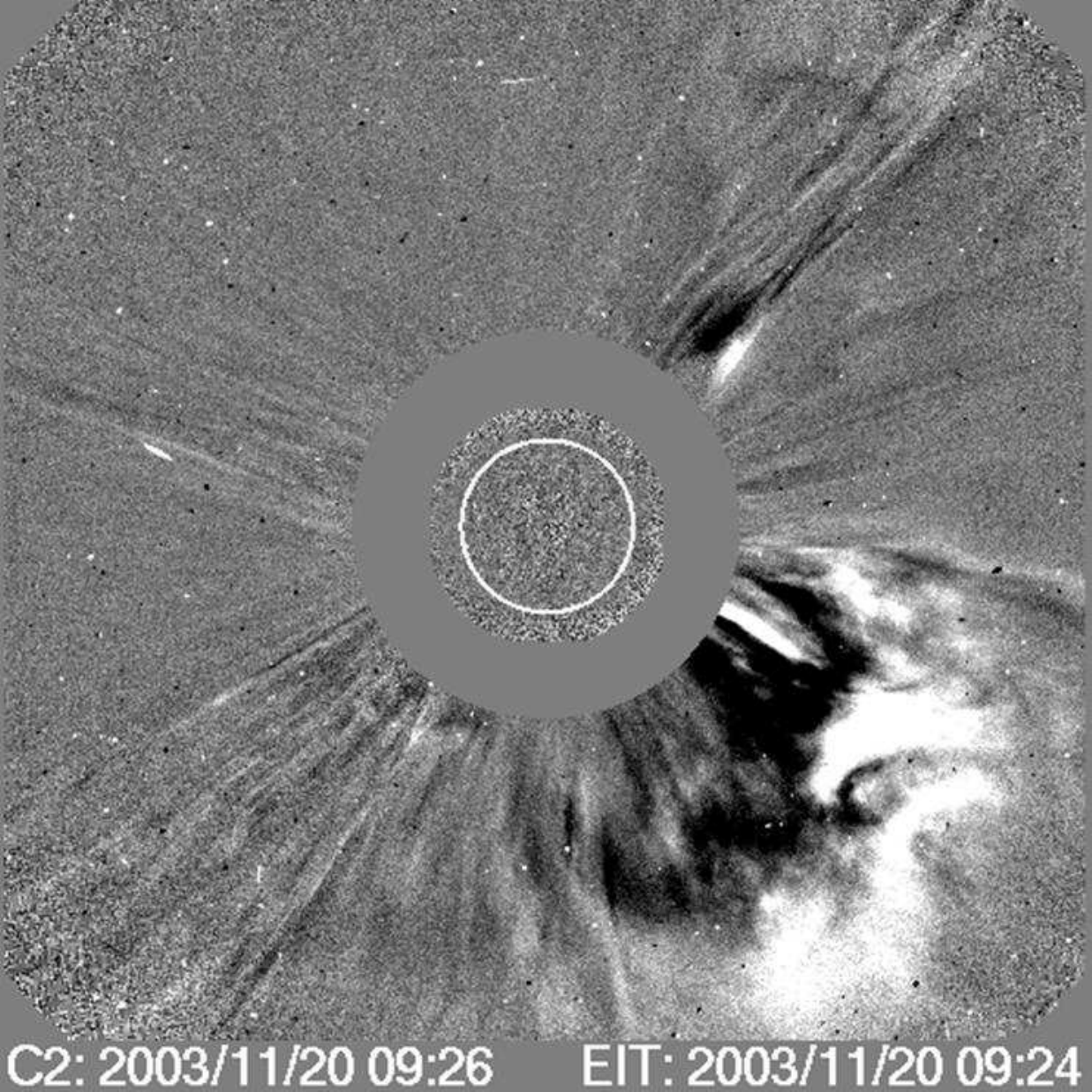}
\includegraphics[width=2.3in]{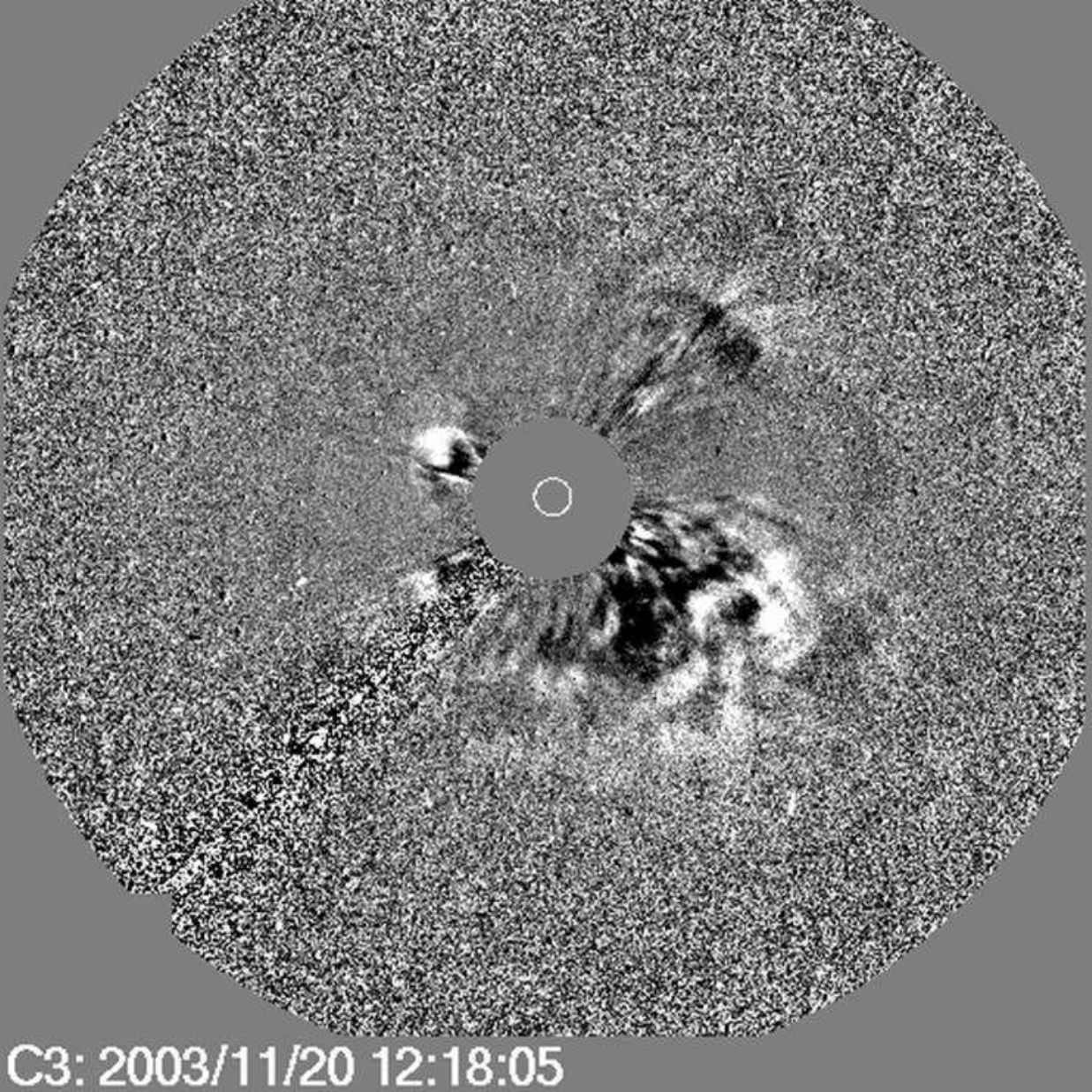}
}}
\caption{Difference images from C2 and C3 LASCO coronagraphs for both CMEs.}
\label{Fig 14}
\end{figure*}

\begin{figure*}
\epsscale{2.0}
\plotone{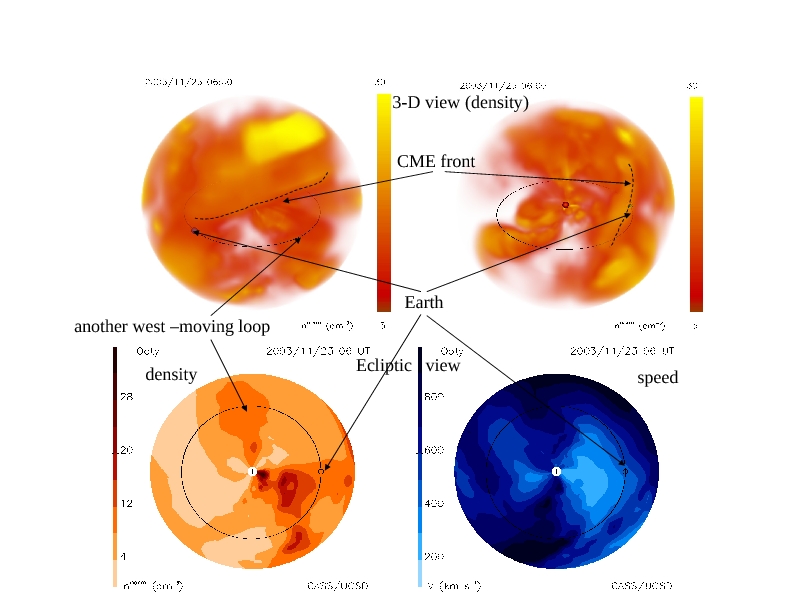}
\caption{3-D view of of the heliosphere obtained from a large number of IPS measurements (i.e., Manoharan 2006). The CME location with respect to the ecliptic plane is shown. The top images are 3-D view and bottom images are ecliptic view.
\label{fig15}}
\end{figure*}
\begin{figure*}
\epsscale{1.5}
\plotone{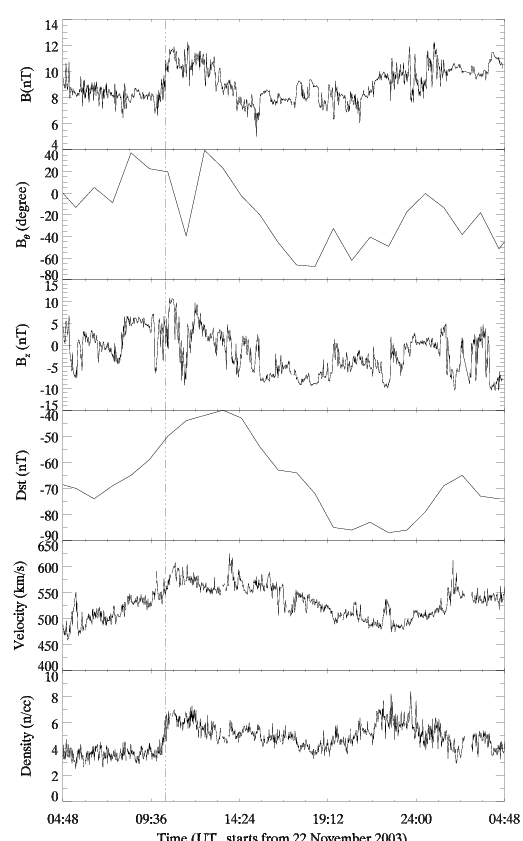}
\caption{The interplanetary observations of magnetic field strength B, B$_\theta$, southward component of magnetic field (B$_z$), geomagnetic index (Dst), solar wind speed and proton density. The arrival of the shock is marked by the vertical line. 
\label{fig16}}
\end{figure*}






\end{document}